\documentclass[%
preprint,
 amsmath,amssymb,
 aps,
]{revtex4-1}
\usepackage[dvipdfmx]{graphicx}
\usepackage{braket}
\usepackage{bm}
\usepackage{amsmath}
\usepackage{dcolumn}

\begin{document}
\title{Monopole and dipole transitions of the cluster states of $^{\bf 18}$O} 
\author{T. Baba}
\affiliation{Kitami Institute of Technology, 090-8507 Kitami, Japan}
\author{M. Kimura}
\affiliation{Department of Physics, Hokkaido University, 060-0810 Sapporo, Japan}
\affiliation{Reaction Nuclear Data Centre (JCPRG),  Hokkaido University, 060-0810
Sapporo, Japan}
\affiliation{Research Center for Nuclear Physics (RCNP), Osaka University, 567-0047 Ibaraki,
Japan} 
\date{\today}

\begin{abstract}
On the basis of an extended antisymmetrized molecular dynamics calculation, we study the cluster
structure and the  of the $0^+$ and $1^-$ states of $^{18}{\rm O}$. We 
discuss that several different kinds of the cluster states appear in the excitation spectrum, and
their monopole and dipole transitions are interesting fingerprints of unique cluster structure. 
We show that the  monopole/dipole transitions are  enhanced for the 
$^{14}{\rm C}$+$\alpha$ cluster states, while they are hindered for the molecular-orbit state. We
also  point out that the ratio of the electric and isoscalar monopole transition strengths gives a
good  hint for the structure of the excited states.  
\end{abstract}

\maketitle
\section{introduction}
The nucleus $^{18}$O has been an important testing ground for our understanding of the clustering
in $N\neq Z$ nuclei. It is of importance and interest to investigate how the extra neutrons affect
and enrich the clustering, since the core nucleus $^{16}$O has the famous ${}^{12}{\rm C}+\alpha$
cluster states
\cite{Arima1967,Horiuchi1968,Buck1975,Suzuki1976,Suzuki1976a,Fujiwara1980,Descouvemont1984}. 

A number of experimental
\cite{Cunsolo1981,Alhassid1982,Gai1983,Gai1987,Gai1989,Curtis2002,Yildiz2006,Johnson2009,Oertzen2009,Avila2014,Yang2019} and
theoretical
\cite{Sakuda1977,Sakuda1978,Assenbaum1984,Baye1984,Descouvemont1985,Suzuki1985,Furutachi2008,Baba2019}
studies have ever explored the  
$^{14}{\rm C}+\alpha$ cluster states in the spectrum of $^{18}$O. They firmly established 
the positive-parity band built on the $0^+_2$ state at 3.63  MeV, which has a 
${}^{14}{\rm C}(0^+_1)+\alpha$ cluster structure
\cite{Cunsolo1981,Sakuda1977,Sakuda1978,Descouvemont1985,Suzuki1985,Furutachi2008,Baba2019}. 
Due to the parity asymmetry of the 
$^{14}{\rm C}+\alpha$ configuration, this band should be accompanied by the negative-parity band
(parity doublet). However, the assignment of the negative-parity band has been rather 
controversial and unsettled.  Gai {\it et al.} \cite{Alhassid1982,Gai1983} assigned the $1^-_1$
state at 4.46 MeV as the doublet partner of the $0^+_2$ state based on the  enhanced $E1$
transition strength between them. However, this assignment was not supported by the theoretical
calculations \cite{Descouvemont1985,Suzuki1985,Furutachi2008,Baba2019}. For example, from the
multi-configuration cluster model calculations, Descouvemont and Baye \cite{Descouvemont1985}
pointed out that the 4.46 MeV state is predominated by the  
$^{14}{\rm C}(2^+)+\alpha$ channel, and hence, cannot be considered as the partner of the
$0^+_2$ state. Alternatively, they showed that the calculated $1^-_3$ state has the pronounced
$^{14}{\rm C}(0^+_1)+\alpha$ cluster structure, and tentatively assigned it to the $1^-$
state observed at 7.62 MeV \cite{Cunsolo1981,Ajzenberg-Selove1983}. Later, this assignment was
corrected by several experiments \cite{Curtis2002,Yildiz2006} which assigned a new negative-parity
band built on the $1^-$ state at 8.03 MeV as the partner of the $0^+_2$ state. A confusing fact is
that this assignment was  again denied by another recent experiment: Avila  {\it et al.}
\cite{Avila2014} reported that  the 
$\alpha$ spectroscopic factor of the 8.03 MeV state is not large, and hence, the state cannot be
a $^{14}{\rm C}(0^+_1)+\alpha$ cluster state. 

In addition to the $^{14}{\rm C}+\alpha$ cluster states, von Oertzen {\it et al.}
\cite{Oertzen2009} proposed a novel type of cluster states which are composed of the 
$^{12}{\rm C}+\alpha$ cluster core and two valence neutrons occupying so-called molecular
orbits (MO state). They tentatively assigned the 7.80 and 10.59 MeV states as the $0^+$ and
$1^-$ doublet of the MO states. The existence of such MO state was qualitatively supported by the
antisymmetrized molecular dynamics (AMD) calculations \cite{Baba2019}, and
experimental efforts to find more convincing evidence is now on going \cite{Yang2019}. Thus,
identifying the pair of the 
$0^+$ and $1^-$ cluster states is important to understand a rich variety of clustering systematics
in $^{18}{\rm O}$.

In this decade, the isoscalar monopole and dipole transition strengths are attracting a lot of
interest as a novel probe for the $0^+$ and $1^-$ cluster states, and have already been used
for the discussion of the clustering in many stable and unstable nuclei 
\cite{Kawabata2007,Kanada-Enyo2007,Funaki2008,Yamada2008,Ito2011,Ichikawa2012,Yamada2012,Kanada-EnYo2014,Yang2014,Chiba2015,Yamada2015,Chiba2016,Kanada-EnYo2016a,Zhou2016,Chiba2017b,Kanada-EnYo2019,Chiba2020,Kanada-EnYo2020}. Therefore,
we expect that they provide a new insight to the clustering of $^{18}{\rm O}$. 

For this purpose, we perform an extended AMD calculation for $^{18}{\rm O}$ taking into account the coupling of
the  $^{14}{\rm C}(0^+_1)+\alpha$ and $^{14}{\rm C}(2^+_1)+\alpha$ channels. We analyze the
cluster structure of the $0^+$ and $1^-$  states referring their $\alpha$-spectroscopic factors,
and investigate how the clustering affects the monopole and dipole transition strengths. We 
show that the dipole transition strength between the doublet of the  $^{14}{\rm C}+\alpha$
cluster states is greatly enhanced, while that of the MO state is hindered. We also discuss that
the ratio of the electric and isoscalar monopole transition strengths also gives us an interesting
hints on the cluster structure.   

This paper is organized as follows: In the next section, we briefly explain how we calculated the
wave functions of the cluster states of $^{18}{\rm O}$. We also explain the  electric and
isoscalar monopole/dipole transition matrices. In the section III, we first review the calculated
and observed spectrum of the cluster states. And we investigate how the characteristics of each
cluster states is reflected to the pattern of the transition strengths. The final section
summarizes this work.

\section{Theoretical framework}
In this study, we use the Hamiltonian same with our previous study \cite{Baba2019},
\begin{align}
 H = \sum_{i=1}^A t_i  - t_{\rm c.m.} + \sum_{i<j}^A v_{ij},
\end{align}
where $t_i$ and $t_{\rm c.m.}$ represents the single-particle and center-of-mass kinetic
energies. $v_{ij}$ includes the Gogny D1S effective nucleon-nucleon interaction \cite{Berger1991}
and Coulomb interaction. 

The model wave function is a parity-projected Slater determinant,
\begin{align}
 \Phi_{\rm AMD} = P^\pi\mathcal{A}\set{\varphi_1,...,\varphi_A},\quad \pi=\pm,
\end{align}
where $P^\pi$ is the parity-projection operator, and the single-particle wave packets has the
deformed Gaussian form \cite{Kanada-Enyo2003,Kimura2004a,Kanada-Enyo2012}, 
\begin{align}
 \varphi_i(\bm r) =& \prod_{\sigma=x,y,z}
 \exp\set{-\nu_\sigma (r_\sigma - Z_{i\sigma})^2}\nonumber\\
 &\otimes\left(\alpha_i\ket{\uparrow} + \beta_i\ket\downarrow\right)
 \otimes\left(\ket{p} {\rm or} \ket{n}\right).
\end{align}
Each Gaussian wave packet has the variational parameters: The Gaussian centroid vector $\bm Z_i$
and the spin parameters $\alpha_i$ and $\beta_i$. The isospin is fixed to either of proton or
neutron. The Gaussian width parameters $\nu_x$, $\nu_y$ and $\nu_z$ are also the variational
parameters and  common to all wave packets.  

These variational parameters are determined by the following two methods. The first is the energy
variation with the constraint, which was already used in our previous study \cite{Baba2019}. Using
the frictional cooling method, the variational parameters are so chosen to 
minimize the total energy under the constraint on the quadruple deformation parameter $\beta$. By
this calculation, we obtain the optimized wave function $\Phi_{\rm AMD}(\beta)$ for each value of 
$\beta$ ($\beta=0.00,0.05,...,1.40$). As discussed in Ref. \cite{Baba2019}, if $\beta$ is small, we
obtain the almost spherical shell-model-like wave functions corresponding to the ground state, but
as $\beta$ increases, we obtain a variety of cluster configurations such as the 
$^{14}{\rm C}+\alpha$ cluster, molecular-orbit state and linear-chain of $\alpha$ particles.

In this study, we extend the model space by applying the second method. We use the Brink-type
wave function for $^{14}{\rm C}+\alpha$ configurations, in which $^{14}{\rm C}$ and $\alpha$
cluster wave functions are placed on the $z$-axis with the inter-cluster distance $d$,
\begin{align}
\Phi_{\rm Brink}(d) = P^\pi\mathcal{A}
 \Set{\Phi_{\alpha}\left(-\frac{14}{18}d\right)\Phi_{\rm C}\left(\frac{4}{18}d\right)},
 \label{eq:brink}
\end{align}
where $\Phi_{\alpha}$ and  $\Phi_{\rm C}$ represents the intrinsic wave functions of 
$^{4}{\rm He}$ and $^{14}{\rm C}$, respectively. $\Phi_{\alpha}$  is assumed to have the $(0s)^4$
configuration, and $\Phi_{\rm C}$ is approximated by a single AMD wave function which has 90\%
overlap with the full  GCM wave function obtained in Ref. \cite{Baba2016,Baba2017}. The
inter-cluster distance $d$ ranges from 0.4 fm to 8.0 fm with the intervals of 0.4 fm.  
Since $\Phi_{\rm C}$ is oblately deformed, we consider three different orientations of $^{14}$C
cluster. 
\begin{figure}[h]
 \centering
 \includegraphics[width=1.0\hsize]{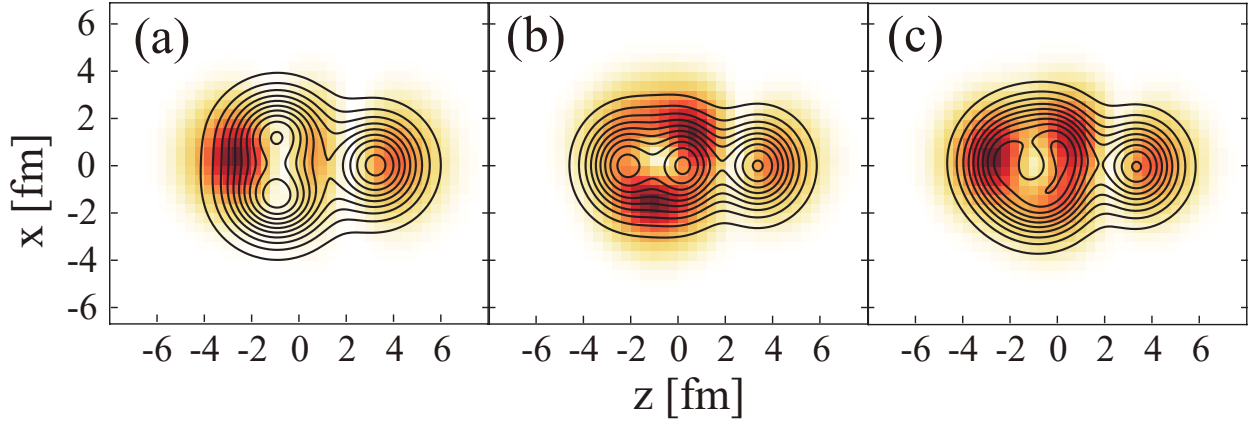}
 \caption{(color online) Density distributions of the Brink-type wave functions in which the
 symmetry axis of the oblately deformed $^{14}{\rm C}$ is directed to the (a) $z$-, (b) $x$-, and
 (c) $y$-axis. The  inter-cluster distance $d$ is fixed to 4.8 fm in all panels.}  
 \label{fig:dens_brk}
\end{figure}
Figure \ref{fig:dens_brk} shows the example of the $^{14}{\rm C}+\alpha$ Brink wave functions
in which the symmetry axis of $^{14}$C is directed to the (a) $z$-, (b) $x$-, and (c)
$y$-axis. Note that the superposition of different orientations of $\Phi_{\rm C}$ naturally handles
the coupling of the $^{14}{\rm C}(0^+_1)+\alpha$ and $^{14}{\rm C}(2^+_1)+\alpha$ channels, which
is known to be important in describing the $^{14}{\rm C}+\alpha$ clustering
\cite{Descouvemont1985}. This point is an advantage of the present calculation compared to
previous AMD studies.  

These wave functions are projected to the eigenstate of the angular momentum and are
superposed to describe the ground and excited states,
\begin{align}
 \Psi^{J^\pi}_{Mp} =& \sum_{Ki}f_{Kip}P^J_{MK}\Phi^\pi_{\rm AMD}(\beta_i)\nonumber\\ 
 &+\sum_{Ki}g_{Kip}P^J_{MK}\Phi^\pi_{\rm Brink}(d_i),\label{eq:gcmwf}
\end{align}
where $P^J_{MK}$ and the index $p$ denote the angular momentum projection operator and the
quantum number other than the angular momentum, respectively. The coefficients $f_{Kip}$
and $g_{Kip}$ are determined by diagonalizing Hamiltonian \cite{Hill1953}.

As a measure of the $^{14}{\rm C}+\alpha$ clustering, we calculate the $\alpha$-spectroscopic
factor. We first calculate the 
$\alpha$ reduced width amplitudes  (RWA) which is the probability amplitude to find the $^{14}{\rm
C}$ and $\alpha$ clusters at the inter-cluster distance $a$. It is defined as the overlap between
the reference cluster state and the wave function given by Eq. (\ref{eq:gcmwf}), 
\begin{align}
y_{j\ell J}(a) &= \sqrt{\frac{18!}{14! 4!}}
 \Braket{\frac{\delta(r-a)}{r^2}\Phi_{\alpha}[\Phi^j_{{\rm C}}Y_{\ell}(\hat r)]^J_M|
  \Psi^{J\pi}_{Mp}},\nonumber\\
 (j&= 0^+ \textit{ \rm  or } 2^+).\label{eq:rwa1}
\end{align}
The reference cluster state (bra state) is the $^{14}{\rm C}(j)+\alpha$ cluster state in which
the $^{14}{\rm C}(j)$ and $\alpha$ clusters are mutually orbiting with inter-cluster distance
$a$, and the intrinsic angular momentum $j$ of the $^{14}{\rm C}$ cluster is coupled with the
orbital angular momentum $\ell$ to the total angular momentum $J$. Here, the $\alpha$ cluster wave
function $\Phi_{\alpha}$ is same with that appears in Eq. (\ref{eq:brink}), and 
the $^{14}{\rm C}$ cluster wave function $\Phi^j_{\rm C}$ also uses the intrinsic wave function
$\Phi_{\rm C}$ same with that appears in Eq. (\ref{eq:brink}) but projected to the eigenstate of
the angular momentum $j^\pi=0^+$ or $2^+$. In the practical calculation, Eq.~(\ref{eq:rwa1})
is evaluated by the Laplace expansion method \cite{Chiba2017}. The $\alpha$ spectroscopic factor  
$S_\alpha$ is given by a squared integral of $y_{j\ell J}$. 
\begin{align}
 S_{\alpha} = \int_0^\infty da\ a^2 |y_{j\ell J}(a)|^2.
\end{align}

In this work, we focus on the electric and isoscalar monopole ($E0$ and $IS0$) and dipole ($E1$
and $IS1$) transition strengths. As  discussed in Refs. \cite{Alhassid1982,Gai1983}, the $E1$
transition strength is a good probe for the $^{14}{\rm C}+\alpha$ cluster states, since the
intrinsic structure  has static dipole moment. In addition to the $E1$ transition strength, in
this decade, the $E0$,  $IS0$ and $IS1$ transition strengths have been regarded and utilized as a
novel probe for the various kinds of clustering \cite{Yamada2008,Chiba2016}, as it was proved that
these transitions from the ground state to an excited cluster state must be considerably
enhanced. The transition operators are defined as follows,    
\begin{align}
 \mathcal{M}^{E0} &= \sum^Z_{i=1} er'^{2}_i, \quad
 \mathcal{M}^{IS0} = \sum^A_{i=1} r'^{2}_i,\\
 \mathcal{M}^{E1}_\mu &= \sum^Z_{i=1} er'_iY_{1\mu}(\hat{r}'_i), \quad
 \mathcal{M}^{IS1}_\mu = \sum^A_{i=1} r'^3_iY_{1\mu}(\hat{r}'_i).
 \end{align}
Note that the single-particle coordinate $\bm r'_i$ is measured from the center-of-mass
$\bm r_{\rm c.m.}$, {\it i.e.}  $\bm r'_i\equiv \bm r_i - \bm r_{\rm c.m.}$, and hence, our
calculation is free from the spurious center-of-mass contributions. The transition strength from
the initial state $\Psi_{0p}^{J^\pi}$ to the $0^+$ state is 
evaluated by the reduced transition matrix,
\begin{align}
 M(\lambda;J^\pi\rightarrow 0^+)=
 \braket{\Psi^{0^+}|\mathcal{M}^\lambda_0|\Psi^{J^\pi}_{0p}} ,
\end{align}
where $\mathcal {M}^\lambda_{0}$ is any of the transition operators where $\lambda$ is either of
$E0$,  $IS0$,  $E1$ and $IS1$.

\section{Results}
\subsection{Cluster states and their structure}
\begin{figure*}[ht!b]
 \centering
 \includegraphics[width=0.9\hsize]{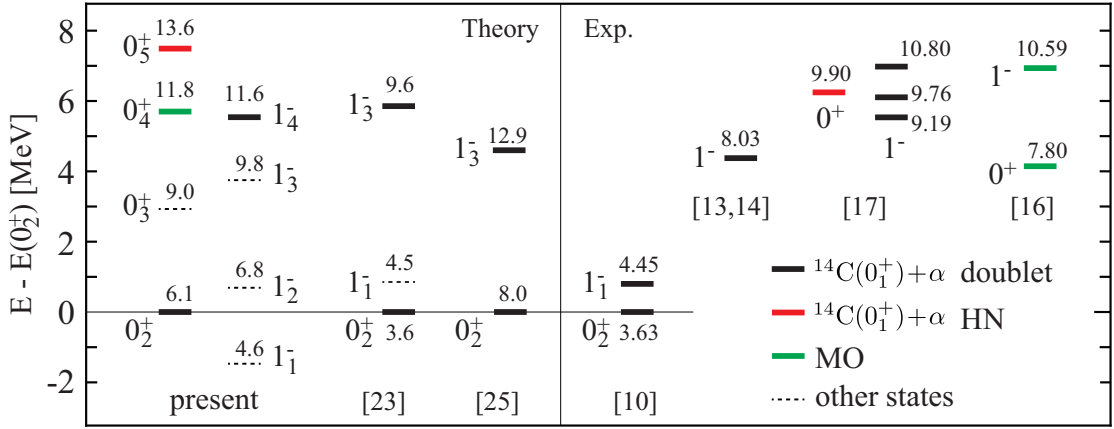}
 \caption{(color online) Spectrum of the $0^+$ and $1^-$ states obtained by the present
 calculation in  comparison with other model calculations
 \cite{Descouvemont1985,Furutachi2008}. Vertical axis shows the 
 energy relative to the  $0^+_2$ state, while numbers in the figure show the energies
 relative to the ground state. The proposed assignments of the cluster states $1^-$ ($0^+$)
 with cluster configurations based on the experiments 
\cite{Cunsolo1981,Ajzenberg-Selove1983,Gai1983,Curtis2002,Yildiz2006,Oertzen2009,Avila2014}
 are also  shown.}
 \label{fig:level}  
\end{figure*}

\begin{figure}[h]
 \centering
 \includegraphics[width=0.9\hsize]{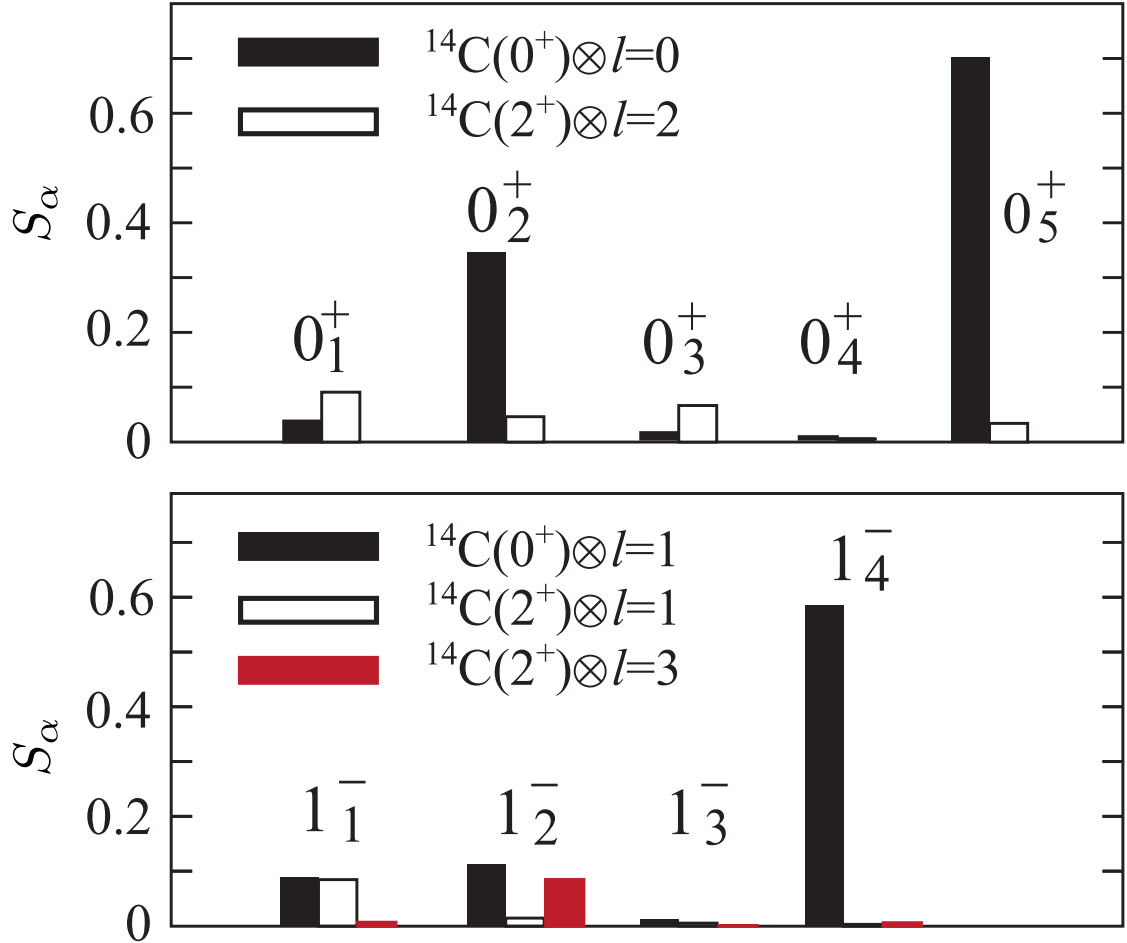}
 \caption{(color online) Calculated $\alpha$ spectroscopic factors of the $0^+$ and $1^-$
 states. }  \label{fig:sfac}
\end{figure}

\begin{figure}[h]
 \centering
 \includegraphics[width=1.0\hsize]{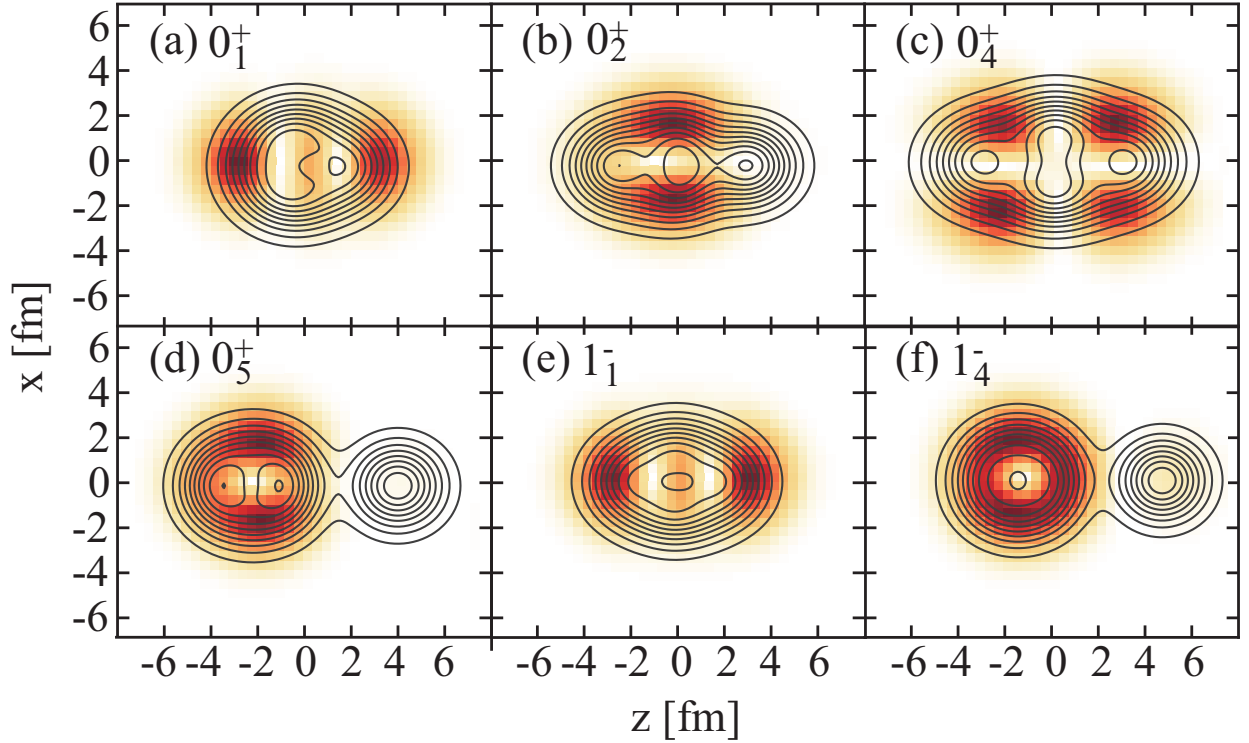}
 \caption{(color online) Density distributions of the intrinsic states which are the dominant
 component of each $0^+$ or $1^-$ state. Solid lines show the proton density distribution, while
 the color plot show that of of two valence neutrons.}  
 \label{fig:dens}
\end{figure}

Figure \ref{fig:level} shows the spectrum of the $0^+$ and $1^-$ states obtained by the present
calculation compared with the other theoretical calculations \cite{Descouvemont1985,Furutachi2008}. It
also shows the  proposed assignments of the cluster states based on the experiments
\cite{Cunsolo1981,Ajzenberg-Selove1983,Gai1983,Curtis2002,Yildiz2006,Oertzen2009,Avila2014}. In the 
present calculation, the bandhead of the positive-parity $^{14}{\rm C}(0^+_1)+\alpha$
cluster band is obtained as the $0^+_2$ state at 6.1 MeV which slightly overestimates the observed 
excitation energy (3.63 MeV). From the calculated spectroscopic factor shown in
Fig. \ref{fig:sfac},  this assignment is rather unique as only this state is dominated by the
$^{14}{\rm C}(0^+_1)+\alpha$ channel in the low-energy region. The intrinsic density of this state
shown in Fig. \ref{fig:dens} (b) also shows a moderate $^{14}{\rm C}+\alpha$ clustering which
is clearly different from the ground state (Fig.\ref{fig:dens} (a)). 

The corresponding negative-parity partner is also uniquely identified as the $1^-_4$ state at 11.6
MeV. It is clear that other $1^-$ states have relatively small spectroscopic
factors and are excluded from the doublet partner. For example, the intrinsic density of the
$1^-_1$ state (Fig. \ref{fig:dens} (e)) clearly shows the absence of the prominent $\alpha$
clustering in this state. Thus, the present calculation yields 5.5 MeV energy splitting between the  
$0^+$ and $1^-$ doublet of the $^{14}{\rm C}(0^+_1)+\alpha$ configuration, which is larger than
that of the $^{12}{\rm C}+\alpha$ doublet of $^{16}{\rm O}$ (3.5 MeV), and as large as
that of the $^{16}{\rm O}+\alpha$ doublet of $^{20}{\rm Ne}$ (5.9 MeV). This indicates the
distortion of the  $^{14}{\rm C}(0^+_1)+\alpha$ clustering in the positive parity state, which
brings about extra binding energy to the $0^+$ state and enlarges the doublet splitting. 
It is notable that other theoretical calculations also yielded similar  
magnitudes of the splitting: Furutachi {\it et al.} \cite{Furutachi2008} reported the 4.9 MeV
splitting from their AMD calculation which used a different effective interaction and model wave
function from ours. Descouvemont {\it et al.} \cite{Descouvemont1985} reported 6.0 MeV splitting
from their multi-configuration cluster model calculation. Thus, theoretical calculations suggest
the consistent magnitudes of the doublet splitting approximately equal to $5\sim 6$ MeV. 

Experimentally, the assignment of the positive-parity $^{14}{\rm C}(0^+_1)+\alpha$ state is well
established and unique
\cite{Cunsolo1981,Sakuda1977,Sakuda1978,Descouvemont1985,Suzuki1985,Furutachi2008,Baba2019}; it
is the $0^+_2$ state at 3.63 MeV, but the assignment of the  
negative-parity state is controversial. Gai {\it et al.} \cite{Gai1983} proposed the $1^-_1$
state at 4.45 MeV as the partner of the $0^+_2$ state, because the $^{14}{\rm C}+\alpha$
intrinsic structure can naturally explain the observed strong $E1$ transition between them. 
However, this assignment gives very small doublet splitting 0.82 MeV which contradicts to
all theoretical calculations. Furthermore, the present calculation shows the $1^-_1$ state is a
mixture of the small amount of the  $^{14}{\rm C}(0^+_1)+\alpha$ and $^{14}{\rm C}(2^+_1)+\alpha$
components, which is 
consistent with the results by Descouvemont {\it et al.} \cite{Descouvemont1985}, but contradict
to the assignment by Gai {\it et al}. Other candidates of the doublet partner are the the $1^-$
state at 8.03 MeV observed by the breakup reaction \cite{Curtis2002,Yildiz2006} and the 9.19, 9.76 and 10.39
MeV states observed by the resonant scattering \cite{Avila2014}. In these assignments, the magnitudes of 
the doublet splitting are approximately in between 4 to 7 MeV. Thus, all these assignments look
compatible with the theoretical results, but there is no conclusive evidence. 

In addition to the $^{14}{\rm C}+\alpha$ doublet, the present calculation predicts two excited
cluster states; the $0^+$ states at 11.8 MeV ($0^+_4$) and 13.6 MeV  ($0^+_5$) whose intrinsic
densities are shown in Fig. \ref{fig:dens} (c) and (d), respectively. The 11.8 MeV state has the
$^{12}{\rm C}+\alpha$ cluster core surrounded by the valence neutrons occupying so-called molecular
orbit (MO state), which is similar to those known for Be, C  and Ne isotopes
\cite{Itagaki2000,VonOertzen2006,Kimura2007,Baba2014,Baba2016,Li2017}.  We consider that  
this state may correspond to the MO structure suggested by von 
Oertzen who proposed a tentative assignment to the 7.80 MeV state \cite{Oertzen2009}. The $0^+_5$
state at 13.6 MeV is the $^{14}{\rm C}+\alpha$ higher-nodal state (HN state), in which the
relative motion between the $^{14}{\rm C}$ and $\alpha$ clusters is excited as  seen in its
density distribution (Fig. \ref{fig:dens} (d)). The corresponding observed state might be the 9.90
MeV state reported by Avila {\it et al.} \cite{Avila2014}, as it is the only $0^+$ state which has
large  spectroscopic factor in this energy region. 

In short, theoretical calculations predict the doublet of the $0^+$ and $1^-$ states with 
the $^{14}{\rm C}+\alpha$ configuration, but the assignment of the $1^-$ state has not been
established uniquely. The $^{14}{\rm C}+\alpha$ HN and MO states are also suggested by
the experiments and the present calculation.

\subsection{Monopole and dipole transitions to and between the cluster states}
\begin{figure}[h!tb]
 \centering
 \includegraphics[width=0.9\hsize]{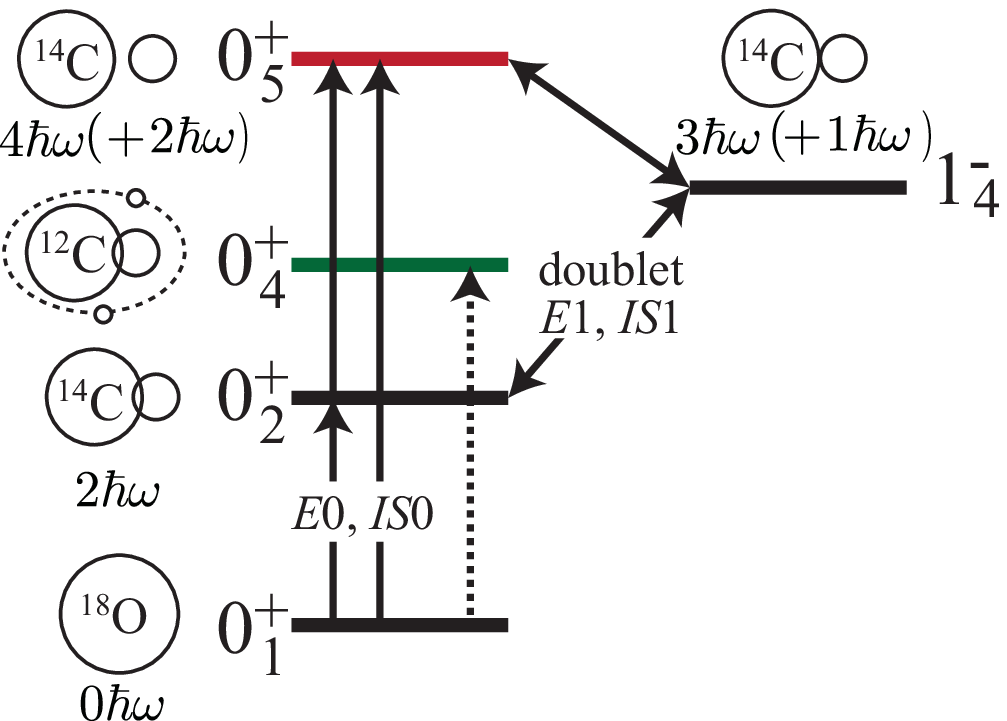}
 \caption{(color online) Schematic figure which illustrates the cluster states of $^{18}{\rm
 O}$  and the pattern of the transitions among them.}   \label{fig:illust}
\end{figure}

Here, we investigate how the characteristics of the cluster states discussed in the previous section
are reflected to the $E0$, $IS0$, $E1$ and $IS1$ transition strengths listed in
Tab. \ref{tab:transition}. For this purpose, the pattern of the calculated transitions
is schematically illustrated in Fig. \ref{fig:illust}. 
\begin{table}[h!]
 \caption{The calculated reduced matrix for the $E0$, $IS0$, $E1$ and $IS1$ transitions in
 Weisskopf unit (in $10^{-2}$ W.u. for the $E1$ transitions). 1 W.u. is equal to 5.93 $e \rm
 fm^2$, 5.93  $\rm fm^2$, 0.665 $e\rm fm$ and 4.39 $\rm fm^3$ for the $E0$, $IS0$, $E1$ and  $IS1$ 
 transitions, respectively.} 
\label{tab:transition}
\begin{center}
 \begin{ruledtabular}
 \begin{tabular}{lD{.}{.}{5}D{.}{.}{5}}
  \multicolumn{1}{l}{$J^\pi_i\rightarrow J^\pi_f$} &
  \multicolumn{1}{c}{$M^{E0}$ [ W.u.]} &
  \multicolumn{1}{c}{$M^{IS0}$ [W.u.]}\\\hline  
  $0^{+}_{1}\rightarrow 0^{+}_{2}$ & 0.36 & 0.67 \\
  $0^{+}_{1}\rightarrow 0^{+}_{3}$ & 0.28 & 0.83\\
  $0^{+}_{1}\rightarrow 0^{+}_{4}$ & 0.02 & 0.02\\
  $0^{+}_{1}\rightarrow 0^{+}_{5}$ & 0.28 & 0.56\\
  $0^{+}_{2}\rightarrow 0^{+}_{5}$ & 1.31 & 2.75\\
  
  \multicolumn{1}{l}{}   &
  \multicolumn{1}{c}{$M^{E1}$ [$10^{-2}$W.u.]} &
  \multicolumn{1}{c}{$M^{IS1}$ [W.u.]} \\ \hline

  $1^{-}_{1}\rightarrow 0^{+}_{1}$ & 1.20 & 0.77\\
  $1^{-}_{2}\rightarrow 0^{+}_{1}$ & 2.53 & 0.51\\
  $1^{-}_{3}\rightarrow 0^{+}_{1}$ & 7.99 & 0.40\\
  $1^{-}_{4}\rightarrow 0^{+}_{1}$ & 3.57 & 0.70\\
  $1^{-}_{1}\rightarrow 0^{+}_{2}$ & 5.11 & 1.58\\
  $1^{-}_{2}\rightarrow 0^{+}_{2}$ & 7.23 & 1.74\\
  $1^{-}_{3}\rightarrow 0^{+}_{2}$ & 2.94 & 0.75\\
  $1^{-}_{4}\rightarrow 0^{+}_{2}$ & 17.3 & 5.19\\
  $1^{-}_{4}\rightarrow 0^{+}_{5}$ & 38.7 & 17.4
 \end{tabular}
 \end{ruledtabular}
\end{center}
\end{table}

In Ref. \cite{Yamada2008}, Yamada {\it al.}  proved that the electric and isoscalar
monopole transitions from the ground state to the $\alpha$-cluster state can be considerably
enhanced. Since the $\alpha$-cluster states appear at relatively small excitation energy compared
to other collective states \cite{Ikeda1968}, the strong $E0$ and $IS0$ strengths at small
excitation energy can be attributed to the $\alpha$-cluster formation. Therefore, the strong
monopole transition has been regarded as a signature of the cluster states in stable and unstable
nuclei
\cite{Kawabata2007,Kanada-Enyo2007,Funaki2008,Yamada2008,Ito2011,Ichikawa2012,Yamada2012,Kanada-EnYo2014,Yang2014,Chiba2015,Yamada2015,Zhou2016,Chiba2020,Kanada-EnYo2020}. In
the present calculation, as expected, we find that the $0^+_2$ state which 
has a  $^{14}{\rm C}+\alpha$ cluster structure has non small $E0$ and $IS0$ transition strengths
from the ground state.  However, they are not as strong as the Weisskopf unit because of the
following reason. As seen in Fig. \ref{fig:sfac} and also discussed in
Ref. \cite{Descouvemont1985}, the dominant cluster component of the ground state is the
$^{14}{\rm C}(2^+_1)+\alpha$ channel, while that of the $0^+_2$ state is different, the $^{14}{\rm
C}(0^+_1)+\alpha$ channel. This mismatch of the internal structure reduces the monopole strengths
between them. The same argument also applies to the $0^+_5$ state.  As the $0^+_5$ state is also
dominated by the $^{14}{\rm C}(0^+_1)+\alpha$ channel, the monopole transition from the ground
state is not so enhanced. On the contrary, the transition between the $0^+_2$ and $0^+_5$ states
is  very strong because they have similar internal structure. It is interesting to note that the
$IS0$ matrices for the $0^+_1\rightarrow 0^+_2$ and $0^+_1\rightarrow 0^+_5$ transitions are
almost twice as large as the $E0$ matrices. This is naturally understood as these excitation are
$\alpha$ clustering, and hence, proton and neutron should equally contribute to the transitions. 
On the other hand,  for the $0^+_1\rightarrow 0^+_3$ transition, the $IS0$ matrix is much larger
than twice of the $E0$ matrix. This is due to the fact that the $0^+_3$ state is not a 
$\alpha$-cluster state but an excited state predominated by the excitation of valence
neutrons. Thus, not only the magnitude but also the ratio of the electric and isoscalar monopole
matrices gives us an insight to nuclear structure. We also note that the $0^+_1\rightarrow 0^+_4$
transitions is rather hindered in both of the electric and isoscalar channels compared to other
states, because the $0^+_4$ has a MO structure, and hence, the $0^+_1\rightarrow 0^+_4$
transition involves the rearrangement of two valence neutrons. A similar hindrance of the monopole
transitions was also discussed in Be isotopes \cite{Ito2011,Yang2014}. 

Experimentally, two different values of $M(E0)$ for the $0^+_1\rightarrow 0^+_2$ transition were
reported: The lower value 0.49 W.u. is not far from our result, but the larger one 1.01 W.u. is
much larger than ours. There is a possibility that we underestimated the $\alpha$ clustering of
the $0^+_2$ state. In particular,  the amount of the $^{14}{\rm C}(2^+)+\alpha$ component in the
$0^+_2$ state may not be large enough in our calculation, as the transition strength is sensitive
to it. We also mention the importance of the rotation effect of $^{14}{\rm C}$ cluster. If we
calculate the $0^+_1\rightarrow 0^+_2$ transition without $^{14}{\rm C}+\alpha$ Brink-basis wave
function, the electric monopole transition strength is 0.18 W.u. which is much smaller than both
of the experimental data. As for the $0^+_1\rightarrow 0^+_{4,5}$ transitions, no corresponding
experimental data has been reported so far. The measurement of the transition strengths from the
ground states to the 7.80 and 9.90 MeV states will provide an interesting hint about the
clustering in $^{18}{\rm O}$, as they are the candidates of the calculated $0^+_{4,5}$ states.  

The electric and isoscalar dipole transitions are good probe to identify the cluster $1^-$ states
as they are enhanced between the doublet ($0^+$ and $1^-$ states) \cite{Alhassid1982,Chiba2016}. 
Indeed, the present results confirm that both of the electric and isoscalar dipole transition
matrices are large for the $1^-_4\rightarrow 0^+_2$ transition. Furthermore, we found that they
are also enhanced for the $1^-_4\rightarrow 0^+_5$  transition, because of the well developed
cluster structure of the  $0^+_5$ state. Experimentally, the 8.03 9.19, 9.76 and 10.39 MeV
states are the candidates of the $1^-_4$ state which constitutes the doublet with the $0^+_2$
state, but the convincing evidence is missing. Therefore, the information about the
magnitude of the dipole transitions of these candidates will be very useful to identify the
doublet.   

Compared to the $1^-_4\rightarrow 0^+_2$ transitions, the the $1^-_4\rightarrow 0^+_1$ transitions
are not so enhanced. This may be again due to the mismatch of the internal structure. The ground
state is dominated by the $^{14}{\rm C}(2^+_1)+\alpha$ channel, while the $1^-_4$ state is not but
by the $^{14}{\rm C}(0^+_1)+\alpha$ channel. We note that a similar  discussion was also made for
the  $1^-_2\rightarrow 0^+_1$ transition of ${}^{16}$O \cite{Kanada-EnYo2019}. Finally, we mention
the $1^-_1\rightarrow 0^+_2$ transitions for which a very strong $E1$ transition was reported by Gai
{\it et al.}  In our calculation, like other theoretical calculations, the enhanced
$E1$ transition was not reproduced, because the $1^-_1$ state exhibits no clustering as seen in its
density plot (Fig. \ref{fig:dens}(e)). This may indicate that some cluster correlations is missing
in theoretical calculations. Thus, unfortunately, the inconsistency between the theory and
experiment for the $1^-_1$ state was not resolved.

\section{Summary}
To understand the clustering systematics in a $N\neq Z$ nucleus $^{18}{\rm O}$, we performed
an extended AMD calculation taking into account the coupling of the $^{14}{\rm C}(0^+_1)+\alpha$
and $^{14}{\rm C}(2^+_1)+\alpha$ channels, and analyzed the cluster structure of the $0^+$ and
$1^-$ states. We also investigated to what extent the characteristics of the cluster states are
reflected to the pattern of the monopole and dipole transition strengths.  

Based on the calculated $\alpha$ spectroscopic factors, we identified the $0^+_2$ and $1^-_4$
states as a doublet of the $^{14}{\rm C}(0^+_1)+\alpha$ cluster states. This assignment gives 5.5
MeV for the doublet splitting which is consistent with other theoretical calculations. 
Furthermore, our calculation predicts the $0^+_4$ and $0^+_5$ states, which respectively have the
MO structure and the $^{14}{\rm C}(0^+_1)+\alpha$ HN structure.  

From the analysis of the electric and isoscalar monopole transitions, we found that the
transitions to the $^{14}{\rm C}+\alpha$ cluster states, namely the $0^+_1\rightarrow 0^+_2$ and
$0^+_1\rightarrow 0^+_5$ transitions, are much stronger than the transition to the MO state
($0^+_1\rightarrow 0^+_4$). This is a good measure to distinguish the  $^{14}{\rm C}+\alpha$
cluster states and MO states, and to identify the experimental counterpart of the $0^+_5$ state.
We also found that for the $0^+_1\rightarrow 0^+_2$ and $0^+_1\rightarrow 0^+_5$ transitions, the
isoscalar transition matrix is approximately twice as large as the electric transition matrix
reflecting the fact that protons and neutrons equally contribute to these $\alpha$ clustering
excitation. 

As for the dipole transitions, we confirmed that the both electric and isoscalar dipole
transitions are greatly enhanced between the $^{14}{\rm C}+\alpha$ cluster states, namely the 
$1^-_4\rightarrow 0^+_2$ and $1^-_4\rightarrow 0^+_5$ transitions. However, we could not resolve
the inconsistency between theories and experiments for the electric dipole transition of the
$1^-_1$  state.

\section{Acknowledgment}
The authors acknowledge the support by the grant for the RCNP joint research project at Osaka
University and by the collaborative research program 2019 at Hokkaido University.
The numerical calculation has been conducted on a supercomputer at Research Center for Nuclear
Physics, Osaka University. One of the author (M.K.) acknowledges the support by the  JSPS KAKENHI
Grant Number JP19K03859.

\bibliography{O18trans.bib}

\begin{thebibliography}{59}%
\makeatletter
\providecommand \@ifxundefined [1]{%
 \@ifx{#1\undefined}
}%
\providecommand \@ifnum [1]{%
 \ifnum #1\expandafter \@firstoftwo
 \else \expandafter \@secondoftwo
 \fi
}%
\providecommand \@ifx [1]{%
 \ifx #1\expandafter \@firstoftwo
 \else \expandafter \@secondoftwo
 \fi
}%
\providecommand \natexlab [1]{#1}%
\providecommand \enquote  [1]{``#1''}%
\providecommand \bibnamefont  [1]{#1}%
\providecommand \bibfnamefont [1]{#1}%
\providecommand \citenamefont [1]{#1}%
\providecommand \href@noop [0]{\@secondoftwo}%
\providecommand \href [0]{\begingroup \@sanitize@url \@href}%
\providecommand \@href[1]{\@@startlink{#1}\@@href}%
\providecommand \@@href[1]{\endgroup#1\@@endlink}%
\providecommand \@sanitize@url [0]{\catcode `\\12\catcode `\$12\catcode
  `\&12\catcode `\#12\catcode `\^12\catcode `\_12\catcode `\%12\relax}%
\providecommand \@@startlink[1]{}%
\providecommand \@@endlink[0]{}%
\providecommand \url  [0]{\begingroup\@sanitize@url \@url }%
\providecommand \@url [1]{\endgroup\@href {#1}{\urlprefix }}%
\providecommand \urlprefix  [0]{URL }%
\providecommand \Eprint [0]{\href }%
\providecommand \doibase [0]{http://dx.doi.org/}%
\providecommand \selectlanguage [0]{\@gobble}%
\providecommand \bibinfo  [0]{\@secondoftwo}%
\providecommand \bibfield  [0]{\@secondoftwo}%
\providecommand \translation [1]{[#1]}%
\providecommand \BibitemOpen [0]{}%
\providecommand \bibitemStop [0]{}%
\providecommand \bibitemNoStop [0]{.\EOS\space}%
\providecommand \EOS [0]{\spacefactor3000\relax}%
\providecommand \BibitemShut  [1]{\csname bibitem#1\endcsname}%
\let\auto@bib@innerbib\@empty
\bibitem [{\citenamefont {Arima}\ \emph {et~al.}(1967)\citenamefont {Arima},
  \citenamefont {Horiuchi},\ and\ \citenamefont {Sebe}}]{Arima1967}%
  \BibitemOpen
  \bibfield  {author} {\bibinfo {author} {\bibfnamefont {A.}~\bibnamefont
  {Arima}}, \bibinfo {author} {\bibfnamefont {H.}~\bibnamefont {Horiuchi}}, \
  and\ \bibinfo {author} {\bibfnamefont {T.}~\bibnamefont {Sebe}},\ }\href
  {\doibase 10.1016/0370-2693(67)90499-6} {\bibfield  {journal} {\bibinfo
  {journal} {Physics Letters B}\ }\textbf {\bibinfo {volume} {24}},\ \bibinfo
  {pages} {129} (\bibinfo {year} {1967})}\BibitemShut {NoStop}%
\bibitem [{\citenamefont {Horiuchi}\ and\ \citenamefont
  {Ikeda}(1968)}]{Horiuchi1968}%
  \BibitemOpen
  \bibfield  {author} {\bibinfo {author} {\bibfnamefont {H.}~\bibnamefont
  {Horiuchi}}\ and\ \bibinfo {author} {\bibfnamefont {K.}~\bibnamefont
  {Ikeda}},\ }\href {\doibase 10.1143/PTP.40.277} {\bibfield  {journal}
  {\bibinfo  {journal} {Progress of Theoretical Physics}\ }\textbf {\bibinfo
  {volume} {40}},\ \bibinfo {pages} {277} (\bibinfo {year} {1968})}\BibitemShut
  {NoStop}%
\bibitem [{\citenamefont {Buck}\ \emph {et~al.}(1975)\citenamefont {Buck},
  \citenamefont {Dover},\ and\ \citenamefont {Vary}}]{Buck1975}%
  \BibitemOpen
  \bibfield  {author} {\bibinfo {author} {\bibfnamefont {B.}~\bibnamefont
  {Buck}}, \bibinfo {author} {\bibfnamefont {C.~B.}\ \bibnamefont {Dover}}, \
  and\ \bibinfo {author} {\bibfnamefont {J.~P.}\ \bibnamefont {Vary}},\ }\href
  {\doibase 10.1103/PhysRevC.11.1803} {\bibfield  {journal} {\bibinfo
  {journal} {Physical Review C}\ }\textbf {\bibinfo {volume} {11}},\ \bibinfo
  {pages} {1803} (\bibinfo {year} {1975})}\BibitemShut {NoStop}%
\bibitem [{\citenamefont {Suzuki}(1976{\natexlab{a}})}]{Suzuki1976}%
  \BibitemOpen
  \bibfield  {author} {\bibinfo {author} {\bibfnamefont {Y.}~\bibnamefont
  {Suzuki}},\ }\href {\doibase 10.1143/ptp.55.1751} {\bibfield  {journal}
  {\bibinfo  {journal} {Progress of Theoretical Physics}\ }\textbf {\bibinfo
  {volume} {55}},\ \bibinfo {pages} {1751} (\bibinfo {year}
  {1976}{\natexlab{a}})}\BibitemShut {NoStop}%
\bibitem [{\citenamefont {Suzuki}(1976{\natexlab{b}})}]{Suzuki1976a}%
  \BibitemOpen
  \bibfield  {author} {\bibinfo {author} {\bibfnamefont {Y.}~\bibnamefont
  {Suzuki}},\ }\href {\doibase 10.1143/ptp.56.111} {\bibfield  {journal}
  {\bibinfo  {journal} {Progress of Theoretical Physics}\ }\textbf {\bibinfo
  {volume} {56}},\ \bibinfo {pages} {111} (\bibinfo {year}
  {1976}{\natexlab{b}})}\BibitemShut {NoStop}%
\bibitem [{\citenamefont {Fujiwara}\ \emph {et~al.}(1980)\citenamefont
  {Fujiwara}, \citenamefont {Horiuchi}, \citenamefont {Ikeda}, \citenamefont
  {Kamimura}, \citenamefont {Kato}, \citenamefont {Suzuki},\ and\ \citenamefont
  {Uegaki}}]{Fujiwara1980}%
  \BibitemOpen
  \bibfield  {author} {\bibinfo {author} {\bibfnamefont {Y.}~\bibnamefont
  {Fujiwara}}, \bibinfo {author} {\bibfnamefont {H.}~\bibnamefont {Horiuchi}},
  \bibinfo {author} {\bibfnamefont {K.}~\bibnamefont {Ikeda}}, \bibinfo
  {author} {\bibfnamefont {M.}~\bibnamefont {Kamimura}}, \bibinfo {author}
  {\bibfnamefont {K.}~\bibnamefont {Kato}}, \bibinfo {author} {\bibfnamefont
  {Y.}~\bibnamefont {Suzuki}}, \ and\ \bibinfo {author} {\bibfnamefont
  {E.}~\bibnamefont {Uegaki}},\ }\href {\doibase 10.1143/PTPS.68.29} {\bibfield
   {journal} {\bibinfo  {journal} {Progress of Theoretical Physics Supplement}\
  }\textbf {\bibinfo {volume} {68}},\ \bibinfo {pages} {29} (\bibinfo {year}
  {1980})}\BibitemShut {NoStop}%
\bibitem [{\citenamefont {Descouvemont}\ \emph {et~al.}(1984)\citenamefont
  {Descouvemont}, \citenamefont {Baye},\ and\ \citenamefont
  {Heenen}}]{Descouvemont1984}%
  \BibitemOpen
  \bibfield  {author} {\bibinfo {author} {\bibfnamefont {P.}~\bibnamefont
  {Descouvemont}}, \bibinfo {author} {\bibfnamefont {D.}~\bibnamefont {Baye}},
  \ and\ \bibinfo {author} {\bibfnamefont {P.~H.}\ \bibnamefont {Heenen}},\
  }\href {\doibase 10.1016/0375-9474(84)90047-2} {\bibfield  {journal}
  {\bibinfo  {journal} {Nuclear Physics, Section A}\ }\textbf {\bibinfo
  {volume} {430}},\ \bibinfo {pages} {426} (\bibinfo {year}
  {1984})}\BibitemShut {NoStop}%
\bibitem [{\citenamefont {Cunsolo}\ \emph {et~al.}(1981)\citenamefont
  {Cunsolo}, \citenamefont {Foti}, \citenamefont {Imm{\`{e}}}, \citenamefont
  {Pappalardo}, \citenamefont {Raciti},\ and\ \citenamefont
  {Saunier}}]{Cunsolo1981}%
  \BibitemOpen
  \bibfield  {author} {\bibinfo {author} {\bibfnamefont {A.}~\bibnamefont
  {Cunsolo}}, \bibinfo {author} {\bibfnamefont {A.}~\bibnamefont {Foti}},
  \bibinfo {author} {\bibfnamefont {G.}~\bibnamefont {Imm{\`{e}}}}, \bibinfo
  {author} {\bibfnamefont {G.}~\bibnamefont {Pappalardo}}, \bibinfo {author}
  {\bibfnamefont {G.}~\bibnamefont {Raciti}}, \ and\ \bibinfo {author}
  {\bibfnamefont {N.}~\bibnamefont {Saunier}},\ }\href {\doibase
  10.1103/PhysRevC.24.476} {\bibfield  {journal} {\bibinfo  {journal} {Physical
  Review C}\ }\textbf {\bibinfo {volume} {24}},\ \bibinfo {pages} {476}
  (\bibinfo {year} {1981})}\BibitemShut {NoStop}%
\bibitem [{\citenamefont {Alhassid}\ \emph {et~al.}(1982)\citenamefont
  {Alhassid}, \citenamefont {Gai},\ and\ \citenamefont
  {Bertsch}}]{Alhassid1982}%
  \BibitemOpen
  \bibfield  {author} {\bibinfo {author} {\bibfnamefont {Y.}~\bibnamefont
  {Alhassid}}, \bibinfo {author} {\bibfnamefont {M.}~\bibnamefont {Gai}}, \
  and\ \bibinfo {author} {\bibfnamefont {G.~F.}\ \bibnamefont {Bertsch}},\
  }\href {\doibase 10.1103/PhysRevLett.49.1482} {\bibfield  {journal} {\bibinfo
   {journal} {Physical Review Letters}\ }\textbf {\bibinfo {volume} {49}},\
  \bibinfo {pages} {1482} (\bibinfo {year} {1982})}\BibitemShut {NoStop}%
\bibitem [{\citenamefont {Gai}\ \emph {et~al.}(1983)\citenamefont {Gai},
  \citenamefont {Ruscev}, \citenamefont {Hayes}, \citenamefont {Ennis},
  \citenamefont {Keddy}, \citenamefont {Schloemer}, \citenamefont {Sterbenz},\
  and\ \citenamefont {Bromley}}]{Gai1983}%
  \BibitemOpen
  \bibfield  {author} {\bibinfo {author} {\bibfnamefont {M.}~\bibnamefont
  {Gai}}, \bibinfo {author} {\bibfnamefont {M.}~\bibnamefont {Ruscev}},
  \bibinfo {author} {\bibfnamefont {A.~C.}\ \bibnamefont {Hayes}}, \bibinfo
  {author} {\bibfnamefont {J.~F.}\ \bibnamefont {Ennis}}, \bibinfo {author}
  {\bibfnamefont {R.}~\bibnamefont {Keddy}}, \bibinfo {author} {\bibfnamefont
  {E.~C.}\ \bibnamefont {Schloemer}}, \bibinfo {author} {\bibfnamefont {S.~M.}\
  \bibnamefont {Sterbenz}}, \ and\ \bibinfo {author} {\bibfnamefont {D.~A.}\
  \bibnamefont {Bromley}},\ }\href {\doibase 10.1103/PhysRevLett.50.239}
  {\bibfield  {journal} {\bibinfo  {journal} {Physical Review Letters}\
  }\textbf {\bibinfo {volume} {50}},\ \bibinfo {pages} {239} (\bibinfo {year}
  {1983})}\BibitemShut {NoStop}%
\bibitem [{\citenamefont {Gai}\ \emph {et~al.}(1987)\citenamefont {Gai},
  \citenamefont {Keddy}, \citenamefont {Bromley}, \citenamefont {Olness},\ and\
  \citenamefont {Warburton}}]{Gai1987}%
  \BibitemOpen
  \bibfield  {author} {\bibinfo {author} {\bibfnamefont {M.}~\bibnamefont
  {Gai}}, \bibinfo {author} {\bibfnamefont {R.}~\bibnamefont {Keddy}}, \bibinfo
  {author} {\bibfnamefont {D.~A.}\ \bibnamefont {Bromley}}, \bibinfo {author}
  {\bibfnamefont {J.~W.}\ \bibnamefont {Olness}}, \ and\ \bibinfo {author}
  {\bibfnamefont {E.~K.}\ \bibnamefont {Warburton}},\ }\href {\doibase
  10.1103/PhysRevC.36.1256} {\bibfield  {journal} {\bibinfo  {journal}
  {Physical Review C}\ }\textbf {\bibinfo {volume} {36}},\ \bibinfo {pages}
  {1256} (\bibinfo {year} {1987})}\BibitemShut {NoStop}%
\bibitem [{\citenamefont {Gai}\ \emph {et~al.}(1989)\citenamefont {Gai},
  \citenamefont {Rugari}, \citenamefont {France}, \citenamefont {Lund},
  \citenamefont {Zhao}, \citenamefont {Bromley}, \citenamefont {Lincoln},
  \citenamefont {Smith}, \citenamefont {Zarcone},\ and\ \citenamefont
  {Kessel}}]{Gai1989}%
  \BibitemOpen
  \bibfield  {author} {\bibinfo {author} {\bibfnamefont {M.}~\bibnamefont
  {Gai}}, \bibinfo {author} {\bibfnamefont {S.~L.}\ \bibnamefont {Rugari}},
  \bibinfo {author} {\bibfnamefont {R.~H.}\ \bibnamefont {France}}, \bibinfo
  {author} {\bibfnamefont {B.~J.}\ \bibnamefont {Lund}}, \bibinfo {author}
  {\bibfnamefont {Z.}~\bibnamefont {Zhao}}, \bibinfo {author} {\bibfnamefont
  {D.~A.}\ \bibnamefont {Bromley}}, \bibinfo {author} {\bibfnamefont {B.~A.}\
  \bibnamefont {Lincoln}}, \bibinfo {author} {\bibfnamefont {W.~W.}\
  \bibnamefont {Smith}}, \bibinfo {author} {\bibfnamefont {M.~J.}\ \bibnamefont
  {Zarcone}}, \ and\ \bibinfo {author} {\bibfnamefont {Q.~C.}\ \bibnamefont
  {Kessel}},\ }\href {\doibase 10.1103/PhysRevLett.62.874} {\bibfield
  {journal} {\bibinfo  {journal} {Physical Review Letters}\ }\textbf {\bibinfo
  {volume} {62}},\ \bibinfo {pages} {874} (\bibinfo {year} {1989})}\BibitemShut
  {NoStop}%
\bibitem [{\citenamefont {Curtis}\ \emph {et~al.}(2002)\citenamefont {Curtis},
  \citenamefont {Caussyn}, \citenamefont {Chandler}, \citenamefont {Cooper},
  \citenamefont {Fletcher}, \citenamefont {Laird},\ and\ \citenamefont
  {Pavan}}]{Curtis2002}%
  \BibitemOpen
  \bibfield  {author} {\bibinfo {author} {\bibfnamefont {N.}~\bibnamefont
  {Curtis}}, \bibinfo {author} {\bibfnamefont {D.~D.}\ \bibnamefont {Caussyn}},
  \bibinfo {author} {\bibfnamefont {C.}~\bibnamefont {Chandler}}, \bibinfo
  {author} {\bibfnamefont {M.~W.}\ \bibnamefont {Cooper}}, \bibinfo {author}
  {\bibfnamefont {N.~R.}\ \bibnamefont {Fletcher}}, \bibinfo {author}
  {\bibfnamefont {R.~W.}\ \bibnamefont {Laird}}, \ and\ \bibinfo {author}
  {\bibfnamefont {J.}~\bibnamefont {Pavan}},\ }\href {\doibase
  10.1103/PhysRevC.66.024315} {\bibfield  {journal} {\bibinfo  {journal}
  {Physical Review C}\ }\textbf {\bibinfo {volume} {66}},\ \bibinfo {pages}
  {024315} (\bibinfo {year} {2002})}\BibitemShut {NoStop}%
\bibitem [{\citenamefont {Yildiz}\ \emph {et~al.}(2006)\citenamefont {Yildiz},
  \citenamefont {Freer}, \citenamefont {Soi{\'{c}}}, \citenamefont {Ahmed},
  \citenamefont {Ashwood}, \citenamefont {Clarke}, \citenamefont {Curtis},
  \citenamefont {Fulton}, \citenamefont {Metelko}, \citenamefont {Novatski},
  \citenamefont {Orr}, \citenamefont {Pitkin}, \citenamefont {Sakuta},\ and\
  \citenamefont {Ziman}}]{Yildiz2006}%
  \BibitemOpen
  \bibfield  {author} {\bibinfo {author} {\bibfnamefont {S.}~\bibnamefont
  {Yildiz}}, \bibinfo {author} {\bibfnamefont {M.}~\bibnamefont {Freer}},
  \bibinfo {author} {\bibfnamefont {N.}~\bibnamefont {Soi{\'{c}}}}, \bibinfo
  {author} {\bibfnamefont {S.}~\bibnamefont {Ahmed}}, \bibinfo {author}
  {\bibfnamefont {N.~I.}\ \bibnamefont {Ashwood}}, \bibinfo {author}
  {\bibfnamefont {N.~M.}\ \bibnamefont {Clarke}}, \bibinfo {author}
  {\bibfnamefont {N.}~\bibnamefont {Curtis}}, \bibinfo {author} {\bibfnamefont
  {B.~R.}\ \bibnamefont {Fulton}}, \bibinfo {author} {\bibfnamefont {C.~J.}\
  \bibnamefont {Metelko}}, \bibinfo {author} {\bibfnamefont {B.}~\bibnamefont
  {Novatski}}, \bibinfo {author} {\bibfnamefont {N.~A.}\ \bibnamefont {Orr}},
  \bibinfo {author} {\bibfnamefont {R.}~\bibnamefont {Pitkin}}, \bibinfo
  {author} {\bibfnamefont {S.}~\bibnamefont {Sakuta}}, \ and\ \bibinfo {author}
  {\bibfnamefont {V.~A.}\ \bibnamefont {Ziman}},\ }\href {\doibase
  10.1103/PhysRevC.73.034601} {\bibfield  {journal} {\bibinfo  {journal}
  {Physical Review C}\ }\textbf {\bibinfo {volume} {73}},\ \bibinfo {pages}
  {034601} (\bibinfo {year} {2006})}\BibitemShut {NoStop}%
\bibitem [{\citenamefont {Johnson}\ \emph {et~al.}(2009)\citenamefont
  {Johnson}, \citenamefont {Rogachev}, \citenamefont {Goldberg}, \citenamefont
  {Brown}, \citenamefont {Robson}, \citenamefont {Crisp}, \citenamefont
  {Cottle}, \citenamefont {Fu}, \citenamefont {Giles}, \citenamefont {Green},
  \citenamefont {Kemper}, \citenamefont {Lee}, \citenamefont {Roeder},\ and\
  \citenamefont {Tribble}}]{Johnson2009}%
  \BibitemOpen
  \bibfield  {author} {\bibinfo {author} {\bibfnamefont {E.~D.}\ \bibnamefont
  {Johnson}}, \bibinfo {author} {\bibfnamefont {G.~V.}\ \bibnamefont
  {Rogachev}}, \bibinfo {author} {\bibfnamefont {V.~Z.}\ \bibnamefont
  {Goldberg}}, \bibinfo {author} {\bibfnamefont {S.}~\bibnamefont {Brown}},
  \bibinfo {author} {\bibfnamefont {D.}~\bibnamefont {Robson}}, \bibinfo
  {author} {\bibfnamefont {A.~M.}\ \bibnamefont {Crisp}}, \bibinfo {author}
  {\bibfnamefont {P.~D.}\ \bibnamefont {Cottle}}, \bibinfo {author}
  {\bibfnamefont {C.}~\bibnamefont {Fu}}, \bibinfo {author} {\bibfnamefont
  {J.}~\bibnamefont {Giles}}, \bibinfo {author} {\bibfnamefont {B.~W.}\
  \bibnamefont {Green}}, \bibinfo {author} {\bibfnamefont {K.~W.}\ \bibnamefont
  {Kemper}}, \bibinfo {author} {\bibfnamefont {K.}~\bibnamefont {Lee}},
  \bibinfo {author} {\bibfnamefont {B.~T.}\ \bibnamefont {Roeder}}, \ and\
  \bibinfo {author} {\bibfnamefont {R.~E.}\ \bibnamefont {Tribble}},\ }\href
  {\doibase 10.1140/EPJA/I2009-10887-1} {\bibfield  {journal} {\bibinfo
  {journal} {The European Physical Journal A 2009 42:2}\ }\textbf {\bibinfo
  {volume} {42}},\ \bibinfo {pages} {135} (\bibinfo {year} {2009})}\BibitemShut
  {NoStop}%
\bibitem [{\citenamefont {von Oertzen}\ \emph {et~al.}(2009)\citenamefont {von
  Oertzen}, \citenamefont {Dorsch}, \citenamefont {Bohlen}, \citenamefont
  {Kr{\"{u}}cken}, \citenamefont {Faestermann}, \citenamefont {Hertenberger},
  \citenamefont {Kokalova}, \citenamefont {Mahgoub}, \citenamefont {Milin},
  \citenamefont {Wheldon},\ and\ \citenamefont {Wirth}}]{Oertzen2009}%
  \BibitemOpen
  \bibfield  {author} {\bibinfo {author} {\bibfnamefont {W.}~\bibnamefont {von
  Oertzen}}, \bibinfo {author} {\bibfnamefont {T.}~\bibnamefont {Dorsch}},
  \bibinfo {author} {\bibfnamefont {H.~G.}\ \bibnamefont {Bohlen}}, \bibinfo
  {author} {\bibfnamefont {R.}~\bibnamefont {Kr{\"{u}}cken}}, \bibinfo {author}
  {\bibfnamefont {T.}~\bibnamefont {Faestermann}}, \bibinfo {author}
  {\bibfnamefont {R.}~\bibnamefont {Hertenberger}}, \bibinfo {author}
  {\bibfnamefont {T.}~\bibnamefont {Kokalova}}, \bibinfo {author}
  {\bibfnamefont {M.}~\bibnamefont {Mahgoub}}, \bibinfo {author} {\bibfnamefont
  {M.}~\bibnamefont {Milin}}, \bibinfo {author} {\bibfnamefont
  {C.}~\bibnamefont {Wheldon}}, \ and\ \bibinfo {author} {\bibfnamefont
  {H.~F.}\ \bibnamefont {Wirth}},\ }\href {\doibase 10.1140/EPJA/I2009-10894-2}
  {\bibfield  {journal} {\bibinfo  {journal} {The European Physical Journal A
  2009 43:1}\ }\textbf {\bibinfo {volume} {43}},\ \bibinfo {pages} {17}
  (\bibinfo {year} {2009})}\BibitemShut {NoStop}%
\bibitem [{\citenamefont {Avila}\ \emph {et~al.}(2014)\citenamefont {Avila},
  \citenamefont {Rogachev}, \citenamefont {Goldberg}, \citenamefont {Johnson},
  \citenamefont {Kemper}, \citenamefont {Tchuvil'sky},\ and\ \citenamefont
  {Volya}}]{Avila2014}%
  \BibitemOpen
  \bibfield  {author} {\bibinfo {author} {\bibfnamefont {M.~L.}\ \bibnamefont
  {Avila}}, \bibinfo {author} {\bibfnamefont {G.~V.}\ \bibnamefont {Rogachev}},
  \bibinfo {author} {\bibfnamefont {V.~Z.}\ \bibnamefont {Goldberg}}, \bibinfo
  {author} {\bibfnamefont {E.~D.}\ \bibnamefont {Johnson}}, \bibinfo {author}
  {\bibfnamefont {K.~W.}\ \bibnamefont {Kemper}}, \bibinfo {author}
  {\bibfnamefont {Y.~M.}\ \bibnamefont {Tchuvil'sky}}, \ and\ \bibinfo {author}
  {\bibfnamefont {A.~S.}\ \bibnamefont {Volya}},\ }\href {\doibase
  10.1103/PhysRevC.90.024327} {\bibfield  {journal} {\bibinfo  {journal}
  {Physical Review C - Nuclear Physics}\ }\textbf {\bibinfo {volume} {90}},\
  \bibinfo {pages} {024327} (\bibinfo {year} {2014})}\BibitemShut {NoStop}%
\bibitem [{\citenamefont {Yang}\ \emph {et~al.}(2019)\citenamefont {Yang},
  \citenamefont {Ye}, \citenamefont {Feng}, \citenamefont {Lin}, \citenamefont
  {Jia}, \citenamefont {Li}, \citenamefont {Lou}, \citenamefont {Li},
  \citenamefont {Ge}, \citenamefont {Yang}, \citenamefont {Hua}, \citenamefont
  {Li}, \citenamefont {Zang}, \citenamefont {Liu}, \citenamefont {Jiang},
  \citenamefont {Li}, \citenamefont {Liu}, \citenamefont {Chen}, \citenamefont
  {Wu}, \citenamefont {Wang}, \citenamefont {Liu}, \citenamefont {Wang},
  \citenamefont {Li}, \citenamefont {Luo}, \citenamefont {Jiang}, \citenamefont
  {Bai}, \citenamefont {Xu}, \citenamefont {Ma}, \citenamefont {Sun},
  \citenamefont {Wang}, \citenamefont {Yang},\ and\ \citenamefont
  {Chen}}]{Yang2019}%
  \BibitemOpen
  \bibfield  {author} {\bibinfo {author} {\bibfnamefont {B.}~\bibnamefont
  {Yang}}, \bibinfo {author} {\bibfnamefont {Y.~L.}\ \bibnamefont {Ye}},
  \bibinfo {author} {\bibfnamefont {J.}~\bibnamefont {Feng}}, \bibinfo {author}
  {\bibfnamefont {C.~J.}\ \bibnamefont {Lin}}, \bibinfo {author} {\bibfnamefont
  {H.~M.}\ \bibnamefont {Jia}}, \bibinfo {author} {\bibfnamefont {Z.~H.}\
  \bibnamefont {Li}}, \bibinfo {author} {\bibfnamefont {J.~L.}\ \bibnamefont
  {Lou}}, \bibinfo {author} {\bibfnamefont {Q.~T.}\ \bibnamefont {Li}},
  \bibinfo {author} {\bibfnamefont {Y.~C.}\ \bibnamefont {Ge}}, \bibinfo
  {author} {\bibfnamefont {X.~F.}\ \bibnamefont {Yang}}, \bibinfo {author}
  {\bibfnamefont {H.}~\bibnamefont {Hua}}, \bibinfo {author} {\bibfnamefont
  {J.}~\bibnamefont {Li}}, \bibinfo {author} {\bibfnamefont {H.~L.}\
  \bibnamefont {Zang}}, \bibinfo {author} {\bibfnamefont {Q.}~\bibnamefont
  {Liu}}, \bibinfo {author} {\bibfnamefont {W.}~\bibnamefont {Jiang}}, \bibinfo
  {author} {\bibfnamefont {C.~G.}\ \bibnamefont {Li}}, \bibinfo {author}
  {\bibfnamefont {Y.}~\bibnamefont {Liu}}, \bibinfo {author} {\bibfnamefont
  {Z.~Q.}\ \bibnamefont {Chen}}, \bibinfo {author} {\bibfnamefont {H.~Y.}\
  \bibnamefont {Wu}}, \bibinfo {author} {\bibfnamefont {C.~G.}\ \bibnamefont
  {Wang}}, \bibinfo {author} {\bibfnamefont {W.}~\bibnamefont {Liu}}, \bibinfo
  {author} {\bibfnamefont {X.}~\bibnamefont {Wang}}, \bibinfo {author}
  {\bibfnamefont {J.~J.}\ \bibnamefont {Li}}, \bibinfo {author} {\bibfnamefont
  {D.~W.}\ \bibnamefont {Luo}}, \bibinfo {author} {\bibfnamefont
  {Y.}~\bibnamefont {Jiang}}, \bibinfo {author} {\bibfnamefont {S.~W.}\
  \bibnamefont {Bai}}, \bibinfo {author} {\bibfnamefont {J.~Y.}\ \bibnamefont
  {Xu}}, \bibinfo {author} {\bibfnamefont {N.~R.}\ \bibnamefont {Ma}}, \bibinfo
  {author} {\bibfnamefont {L.~J.}\ \bibnamefont {Sun}}, \bibinfo {author}
  {\bibfnamefont {D.~X.}\ \bibnamefont {Wang}}, \bibinfo {author}
  {\bibfnamefont {Z.~H.}\ \bibnamefont {Yang}}, \ and\ \bibinfo {author}
  {\bibfnamefont {J.}~\bibnamefont {Chen}},\ }\href {\doibase
  10.1103/PhysRevC.99.064315} {\bibfield  {journal} {\bibinfo  {journal}
  {Physical Review C}\ }\textbf {\bibinfo {volume} {99}},\ \bibinfo {pages}
  {064315} (\bibinfo {year} {2019})}\BibitemShut {NoStop}%
\bibitem [{\citenamefont {Sakuda}(1977)}]{Sakuda1977}%
  \BibitemOpen
  \bibfield  {author} {\bibinfo {author} {\bibfnamefont {T.}~\bibnamefont
  {Sakuda}},\ }\href {\doibase 10.1143/ptp.57.855} {\bibfield  {journal}
  {\bibinfo  {journal} {Progress of Theoretical Physics}\ }\textbf {\bibinfo
  {volume} {57}},\ \bibinfo {pages} {855} (\bibinfo {year} {1977})}\BibitemShut
  {NoStop}%
\bibitem [{\citenamefont {Sakuda}\ \emph {et~al.}(1978)\citenamefont {Sakuda},
  \citenamefont {Nagata},\ and\ \citenamefont {Nemoto}}]{Sakuda1978}%
  \BibitemOpen
  \bibfield  {author} {\bibinfo {author} {\bibfnamefont {T.}~\bibnamefont
  {Sakuda}}, \bibinfo {author} {\bibfnamefont {S.}~\bibnamefont {Nagata}}, \
  and\ \bibinfo {author} {\bibfnamefont {F.}~\bibnamefont {Nemoto}},\ }\href
  {\doibase 10.1143/ptp.59.1543} {\bibfield  {journal} {\bibinfo  {journal}
  {Progress of Theoretical Physics}\ }\textbf {\bibinfo {volume} {59}},\
  \bibinfo {pages} {1543} (\bibinfo {year} {1978})}\BibitemShut {NoStop}%
\bibitem [{\citenamefont {Assenbaum}\ \emph {et~al.}(1984)\citenamefont
  {Assenbaum}, \citenamefont {Langanke},\ and\ \citenamefont
  {Weiguny}}]{Assenbaum1984}%
  \BibitemOpen
  \bibfield  {author} {\bibinfo {author} {\bibfnamefont {H.~J.}\ \bibnamefont
  {Assenbaum}}, \bibinfo {author} {\bibfnamefont {K.}~\bibnamefont {Langanke}},
  \ and\ \bibinfo {author} {\bibfnamefont {A.}~\bibnamefont {Weiguny}},\ }\href
  {\doibase 10.1007/BF02117212} {\bibfield  {journal} {\bibinfo  {journal}
  {Zeitschrift f{\"{u}}r Physik A: Atoms and Nuclei}\ }\textbf {\bibinfo
  {volume} {318}},\ \bibinfo {pages} {35} (\bibinfo {year} {1984})}\BibitemShut
  {NoStop}%
\bibitem [{\citenamefont {Baye}\ and\ \citenamefont
  {Descouvemont}(1984)}]{Baye1984}%
  \BibitemOpen
  \bibfield  {author} {\bibinfo {author} {\bibfnamefont {D.}~\bibnamefont
  {Baye}}\ and\ \bibinfo {author} {\bibfnamefont {P.}~\bibnamefont
  {Descouvemont}},\ }\href {\doibase 10.1016/0370-2693(84)91697-6} {\bibfield
  {journal} {\bibinfo  {journal} {Physics Letters B}\ }\textbf {\bibinfo
  {volume} {146}},\ \bibinfo {pages} {285} (\bibinfo {year}
  {1984})}\BibitemShut {NoStop}%
\bibitem [{\citenamefont {Descouvemont}\ and\ \citenamefont
  {Baye}(1985)}]{Descouvemont1985}%
  \BibitemOpen
  \bibfield  {author} {\bibinfo {author} {\bibfnamefont {P.}~\bibnamefont
  {Descouvemont}}\ and\ \bibinfo {author} {\bibfnamefont {D.}~\bibnamefont
  {Baye}},\ }\href {\doibase 10.1103/PhysRevC.31.2274} {\bibfield  {journal}
  {\bibinfo  {journal} {Physical Review C}\ }\textbf {\bibinfo {volume} {31}},\
  \bibinfo {pages} {2274} (\bibinfo {year} {1985})}\BibitemShut {NoStop}%
\bibitem [{\citenamefont {Suzuki}\ \emph {et~al.}(1985)\citenamefont {Suzuki},
  \citenamefont {Yamamoto},\ and\ \citenamefont {Ikeda}}]{Suzuki1985}%
  \BibitemOpen
  \bibfield  {author} {\bibinfo {author} {\bibfnamefont {Y.}~\bibnamefont
  {Suzuki}}, \bibinfo {author} {\bibfnamefont {A.}~\bibnamefont {Yamamoto}}, \
  and\ \bibinfo {author} {\bibfnamefont {K.}~\bibnamefont {Ikeda}},\ }\href
  {\doibase 10.1016/0375-9474(85)90457-9} {\bibfield  {journal} {\bibinfo
  {journal} {Nuclear Physics, Section A}\ }\textbf {\bibinfo {volume} {444}},\
  \bibinfo {pages} {365} (\bibinfo {year} {1985})}\BibitemShut {NoStop}%
\bibitem [{\citenamefont {Furutachi}\ \emph {et~al.}(2008)\citenamefont
  {Furutachi}, \citenamefont {Kimura}, \citenamefont {Dote}, \citenamefont
  {Kanada-En'yo}, \citenamefont {Oryu}, \citenamefont {Dot{\'{e}}},
  \citenamefont {Kanada-En'yo},\ and\ \citenamefont {Oryu}}]{Furutachi2008}%
  \BibitemOpen
  \bibfield  {author} {\bibinfo {author} {\bibfnamefont {N.}~\bibnamefont
  {Furutachi}}, \bibinfo {author} {\bibfnamefont {M.}~\bibnamefont {Kimura}},
  \bibinfo {author} {\bibfnamefont {A.}~\bibnamefont {Dote}}, \bibinfo {author}
  {\bibfnamefont {Y.}~\bibnamefont {Kanada-En'yo}}, \bibinfo {author}
  {\bibfnamefont {S.}~\bibnamefont {Oryu}}, \bibinfo {author} {\bibfnamefont
  {A.}~\bibnamefont {Dot{\'{e}}}}, \bibinfo {author} {\bibfnamefont
  {Y.}~\bibnamefont {Kanada-En'yo}}, \ and\ \bibinfo {author} {\bibfnamefont
  {S.}~\bibnamefont {Oryu}},\ }\href {\doibase 10.1143/PTP.119.403} {\bibfield
  {journal} {\bibinfo  {journal} {Progress of Theoretical Physics}\ }\textbf
  {\bibinfo {volume} {119}},\ \bibinfo {pages} {403} (\bibinfo {year}
  {2008})}\BibitemShut {NoStop}%
\bibitem [{\citenamefont {Baba}\ and\ \citenamefont {Kimura}(2019)}]{Baba2019}%
  \BibitemOpen
  \bibfield  {author} {\bibinfo {author} {\bibfnamefont {T.}~\bibnamefont
  {Baba}}\ and\ \bibinfo {author} {\bibfnamefont {M.}~\bibnamefont {Kimura}},\
  }\href {\doibase 10.1103/PhysRevC.100.064311} {\bibfield  {journal} {\bibinfo
   {journal} {Physical Review C}\ }\textbf {\bibinfo {volume} {100}},\ \bibinfo
  {pages} {064311} (\bibinfo {year} {2019})}\BibitemShut {NoStop}%
\bibitem [{\citenamefont {Ajzenberg-Selove}(1983)}]{Ajzenberg-Selove1983}%
  \BibitemOpen
  \bibfield  {author} {\bibinfo {author} {\bibfnamefont {F.}~\bibnamefont
  {Ajzenberg-Selove}},\ }\href {\doibase 10.1016/0375-9474(83)90180-X}
  {\bibfield  {journal} {\bibinfo  {journal} {Nuclear Physics, Section A}\
  }\textbf {\bibinfo {volume} {392}},\ \bibinfo {pages} {1} (\bibinfo {year}
  {1983})}\BibitemShut {NoStop}%
\bibitem [{\citenamefont {Kawabata}\ \emph {et~al.}(2007)\citenamefont
  {Kawabata}, \citenamefont {Akimune}, \citenamefont {Fujita}, \citenamefont
  {Fujita}, \citenamefont {Fujiwara}, \citenamefont {Hara}, \citenamefont
  {Hatanaka}, \citenamefont {Itoh}, \citenamefont {Kanada-En'yo}, \citenamefont
  {Kishi}, \citenamefont {Nakanishi}, \citenamefont {Sakaguchi}, \citenamefont
  {Shimbara}, \citenamefont {Tamii}, \citenamefont {Terashima}, \citenamefont
  {Uchida}, \citenamefont {Wakasa}, \citenamefont {Yasuda}, \citenamefont
  {Yoshida},\ and\ \citenamefont {Yosoi}}]{Kawabata2007}%
  \BibitemOpen
  \bibfield  {author} {\bibinfo {author} {\bibfnamefont {T.}~\bibnamefont
  {Kawabata}}, \bibinfo {author} {\bibfnamefont {H.}~\bibnamefont {Akimune}},
  \bibinfo {author} {\bibfnamefont {H.}~\bibnamefont {Fujita}}, \bibinfo
  {author} {\bibfnamefont {Y.}~\bibnamefont {Fujita}}, \bibinfo {author}
  {\bibfnamefont {M.}~\bibnamefont {Fujiwara}}, \bibinfo {author}
  {\bibfnamefont {K.}~\bibnamefont {Hara}}, \bibinfo {author} {\bibfnamefont
  {K.}~\bibnamefont {Hatanaka}}, \bibinfo {author} {\bibfnamefont
  {M.}~\bibnamefont {Itoh}}, \bibinfo {author} {\bibfnamefont {Y.}~\bibnamefont
  {Kanada-En'yo}}, \bibinfo {author} {\bibfnamefont {S.}~\bibnamefont {Kishi}},
  \bibinfo {author} {\bibfnamefont {K.}~\bibnamefont {Nakanishi}}, \bibinfo
  {author} {\bibfnamefont {H.}~\bibnamefont {Sakaguchi}}, \bibinfo {author}
  {\bibfnamefont {Y.}~\bibnamefont {Shimbara}}, \bibinfo {author}
  {\bibfnamefont {A.}~\bibnamefont {Tamii}}, \bibinfo {author} {\bibfnamefont
  {S.}~\bibnamefont {Terashima}}, \bibinfo {author} {\bibfnamefont
  {M.}~\bibnamefont {Uchida}}, \bibinfo {author} {\bibfnamefont
  {T.}~\bibnamefont {Wakasa}}, \bibinfo {author} {\bibfnamefont
  {Y.}~\bibnamefont {Yasuda}}, \bibinfo {author} {\bibfnamefont
  {H.}~\bibnamefont {Yoshida}}, \ and\ \bibinfo {author} {\bibfnamefont
  {M.}~\bibnamefont {Yosoi}},\ }\href {\doibase 10.1016/j.physletb.2006.11.079}
  {\bibfield  {journal} {\bibinfo  {journal} {Physics Letters B}\ }\textbf
  {\bibinfo {volume} {646}},\ \bibinfo {pages} {6} (\bibinfo {year}
  {2007})}\BibitemShut {NoStop}%
\bibitem [{\citenamefont {Kanada-En'yo}(2007)}]{Kanada-Enyo2007}%
  \BibitemOpen
  \bibfield  {author} {\bibinfo {author} {\bibfnamefont {Y.}~\bibnamefont
  {Kanada-En'yo}},\ }\href {\doibase 10.1103/PhysRevC.75.024302} {\bibfield
  {journal} {\bibinfo  {journal} {Physical Review C - Nuclear Physics}\
  }\textbf {\bibinfo {volume} {75}},\ \bibinfo {pages} {024302} (\bibinfo
  {year} {2007})}\BibitemShut {NoStop}%
\bibitem [{\citenamefont {Funaki}\ \emph {et~al.}(2008)\citenamefont {Funaki},
  \citenamefont {Yamada}, \citenamefont {Horiuchi}, \citenamefont
  {R{\"{o}}pke}, \citenamefont {Schuck},\ and\ \citenamefont
  {Tohsaki}}]{Funaki2008}%
  \BibitemOpen
  \bibfield  {author} {\bibinfo {author} {\bibfnamefont {Y.}~\bibnamefont
  {Funaki}}, \bibinfo {author} {\bibfnamefont {T.}~\bibnamefont {Yamada}},
  \bibinfo {author} {\bibfnamefont {H.}~\bibnamefont {Horiuchi}}, \bibinfo
  {author} {\bibfnamefont {G.}~\bibnamefont {R{\"{o}}pke}}, \bibinfo {author}
  {\bibfnamefont {P.}~\bibnamefont {Schuck}}, \ and\ \bibinfo {author}
  {\bibfnamefont {A.}~\bibnamefont {Tohsaki}},\ }\href {\doibase
  10.1103/PhysRevLett.101.082502} {\bibfield  {journal} {\bibinfo  {journal}
  {Physical Review Letters}\ }\textbf {\bibinfo {volume} {101}},\ \bibinfo
  {pages} {082502} (\bibinfo {year} {2008})}\BibitemShut {NoStop}%
\bibitem [{\citenamefont {Yamada}\ \emph {et~al.}(2008)\citenamefont {Yamada},
  \citenamefont {Funaki}, \citenamefont {Horiuchi}, \citenamefont {Ikeda},\
  and\ \citenamefont {Tohsaki}}]{Yamada2008}%
  \BibitemOpen
  \bibfield  {author} {\bibinfo {author} {\bibfnamefont {T.}~\bibnamefont
  {Yamada}}, \bibinfo {author} {\bibfnamefont {Y.}~\bibnamefont {Funaki}},
  \bibinfo {author} {\bibfnamefont {H.}~\bibnamefont {Horiuchi}}, \bibinfo
  {author} {\bibfnamefont {K.}~\bibnamefont {Ikeda}}, \ and\ \bibinfo {author}
  {\bibfnamefont {A.}~\bibnamefont {Tohsaki}},\ }\href {\doibase
  10.1143/PTP.120.1139} {\bibfield  {journal} {\bibinfo  {journal} {Progress of
  Theoretical Physics}\ }\textbf {\bibinfo {volume} {120}},\ \bibinfo {pages}
  {1139} (\bibinfo {year} {2008})}\BibitemShut {NoStop}%
\bibitem [{\citenamefont {Ito}(2011)}]{Ito2011}%
  \BibitemOpen
  \bibfield  {author} {\bibinfo {author} {\bibfnamefont {M.}~\bibnamefont
  {Ito}},\ }\href {\doibase 10.1103/PhysRevC.83.044319} {\bibfield  {journal}
  {\bibinfo  {journal} {Physical Review C - Nuclear Physics}\ }\textbf
  {\bibinfo {volume} {83}},\ \bibinfo {pages} {044319} (\bibinfo {year}
  {2011})}\BibitemShut {NoStop}%
\bibitem [{\citenamefont {Ichikawa}\ \emph {et~al.}(2012)\citenamefont
  {Ichikawa}, \citenamefont {Itagaki}, \citenamefont {Kanada-En'Yo},
  \citenamefont {Kokalova},\ and\ \citenamefont {{Von
  Oertzen}}}]{Ichikawa2012}%
  \BibitemOpen
  \bibfield  {author} {\bibinfo {author} {\bibfnamefont {T.}~\bibnamefont
  {Ichikawa}}, \bibinfo {author} {\bibfnamefont {N.}~\bibnamefont {Itagaki}},
  \bibinfo {author} {\bibfnamefont {Y.}~\bibnamefont {Kanada-En'Yo}}, \bibinfo
  {author} {\bibfnamefont {T.}~\bibnamefont {Kokalova}}, \ and\ \bibinfo
  {author} {\bibfnamefont {W.}~\bibnamefont {{Von Oertzen}}},\ }\href {\doibase
  10.1103/PhysRevC.86.031303} {\bibfield  {journal} {\bibinfo  {journal}
  {Physical Review C - Nuclear Physics}\ }\textbf {\bibinfo {volume} {86}},\
  \bibinfo {pages} {031303} (\bibinfo {year} {2012})}\BibitemShut {NoStop}%
\bibitem [{\citenamefont {Yamada}\ \emph {et~al.}(2012)\citenamefont {Yamada},
  \citenamefont {Funaki}, \citenamefont {Myo}, \citenamefont {Horiuchi},
  \citenamefont {Ikeda}, \citenamefont {R{\"{o}}pke}, \citenamefont {Schuck},\
  and\ \citenamefont {Tohsaki}}]{Yamada2012}%
  \BibitemOpen
  \bibfield  {author} {\bibinfo {author} {\bibfnamefont {T.}~\bibnamefont
  {Yamada}}, \bibinfo {author} {\bibfnamefont {Y.}~\bibnamefont {Funaki}},
  \bibinfo {author} {\bibfnamefont {T.}~\bibnamefont {Myo}}, \bibinfo {author}
  {\bibfnamefont {H.}~\bibnamefont {Horiuchi}}, \bibinfo {author}
  {\bibfnamefont {K.}~\bibnamefont {Ikeda}}, \bibinfo {author} {\bibfnamefont
  {G.}~\bibnamefont {R{\"{o}}pke}}, \bibinfo {author} {\bibfnamefont
  {P.}~\bibnamefont {Schuck}}, \ and\ \bibinfo {author} {\bibfnamefont
  {A.}~\bibnamefont {Tohsaki}},\ }\href {\doibase 10.1103/PhysRevC.85.034315}
  {\bibfield  {journal} {\bibinfo  {journal} {Physical Review C - Nuclear
  Physics}\ }\textbf {\bibinfo {volume} {85}},\ \bibinfo {pages} {034315}
  (\bibinfo {year} {2012})}\BibitemShut {NoStop}%
\bibitem [{\citenamefont {Kanada-En'Yo}(2014)}]{Kanada-EnYo2014}%
  \BibitemOpen
  \bibfield  {author} {\bibinfo {author} {\bibfnamefont {Y.}~\bibnamefont
  {Kanada-En'Yo}},\ }\href {\doibase 10.1103/PhysRevC.89.024302} {\bibfield
  {journal} {\bibinfo  {journal} {Physical Review C - Nuclear Physics}\
  }\textbf {\bibinfo {volume} {89}},\ \bibinfo {pages} {024302} (\bibinfo
  {year} {2014})}\BibitemShut {NoStop}%
\bibitem [{\citenamefont {Yang}\ \emph {et~al.}(2014)\citenamefont {Yang},
  \citenamefont {Ye}, \citenamefont {Li}, \citenamefont {Lou}, \citenamefont
  {Wang}, \citenamefont {Jiang}, \citenamefont {Ge}, \citenamefont {Li},
  \citenamefont {Hua}, \citenamefont {Li}, \citenamefont {Xu}, \citenamefont
  {Pei}, \citenamefont {Qiao}, \citenamefont {You}, \citenamefont {Wang},
  \citenamefont {Tian}, \citenamefont {Li}, \citenamefont {Sun}, \citenamefont
  {Liu}, \citenamefont {Chen}, \citenamefont {Wu}, \citenamefont {Li},
  \citenamefont {Jiang}, \citenamefont {Wen}, \citenamefont {Yang},
  \citenamefont {Yang}, \citenamefont {Ma}, \citenamefont {Ma}, \citenamefont
  {Jin}, \citenamefont {Han},\ and\ \citenamefont {Lee}}]{Yang2014}%
  \BibitemOpen
  \bibfield  {author} {\bibinfo {author} {\bibfnamefont {Z.~H.}\ \bibnamefont
  {Yang}}, \bibinfo {author} {\bibfnamefont {Y.~L.}\ \bibnamefont {Ye}},
  \bibinfo {author} {\bibfnamefont {Z.~H.}\ \bibnamefont {Li}}, \bibinfo
  {author} {\bibfnamefont {J.~L.}\ \bibnamefont {Lou}}, \bibinfo {author}
  {\bibfnamefont {J.~S.}\ \bibnamefont {Wang}}, \bibinfo {author}
  {\bibfnamefont {D.~X.}\ \bibnamefont {Jiang}}, \bibinfo {author}
  {\bibfnamefont {Y.~C.}\ \bibnamefont {Ge}}, \bibinfo {author} {\bibfnamefont
  {Q.~T.}\ \bibnamefont {Li}}, \bibinfo {author} {\bibfnamefont
  {H.}~\bibnamefont {Hua}}, \bibinfo {author} {\bibfnamefont {X.~Q.}\
  \bibnamefont {Li}}, \bibinfo {author} {\bibfnamefont {F.~R.}\ \bibnamefont
  {Xu}}, \bibinfo {author} {\bibfnamefont {J.~C.}\ \bibnamefont {Pei}},
  \bibinfo {author} {\bibfnamefont {R.}~\bibnamefont {Qiao}}, \bibinfo {author}
  {\bibfnamefont {H.~B.}\ \bibnamefont {You}}, \bibinfo {author} {\bibfnamefont
  {H.}~\bibnamefont {Wang}}, \bibinfo {author} {\bibfnamefont {Z.~Y.}\
  \bibnamefont {Tian}}, \bibinfo {author} {\bibfnamefont {K.~A.}\ \bibnamefont
  {Li}}, \bibinfo {author} {\bibfnamefont {Y.~L.}\ \bibnamefont {Sun}},
  \bibinfo {author} {\bibfnamefont {H.~N.}\ \bibnamefont {Liu}}, \bibinfo
  {author} {\bibfnamefont {J.}~\bibnamefont {Chen}}, \bibinfo {author}
  {\bibfnamefont {J.}~\bibnamefont {Wu}}, \bibinfo {author} {\bibfnamefont
  {J.}~\bibnamefont {Li}}, \bibinfo {author} {\bibfnamefont {W.}~\bibnamefont
  {Jiang}}, \bibinfo {author} {\bibfnamefont {C.}~\bibnamefont {Wen}}, \bibinfo
  {author} {\bibfnamefont {B.}~\bibnamefont {Yang}}, \bibinfo {author}
  {\bibfnamefont {Y.~Y.}\ \bibnamefont {Yang}}, \bibinfo {author}
  {\bibfnamefont {P.}~\bibnamefont {Ma}}, \bibinfo {author} {\bibfnamefont
  {J.~B.}\ \bibnamefont {Ma}}, \bibinfo {author} {\bibfnamefont {S.~L.}\
  \bibnamefont {Jin}}, \bibinfo {author} {\bibfnamefont {J.~L.}\ \bibnamefont
  {Han}}, \ and\ \bibinfo {author} {\bibfnamefont {J.}~\bibnamefont {Lee}},\
  }\href {\doibase 10.1103/PhysRevLett.112.162501} {\bibfield  {journal}
  {\bibinfo  {journal} {Physical Review Letters}\ }\textbf {\bibinfo {volume}
  {112}},\ \bibinfo {pages} {162501} (\bibinfo {year} {2014})}\BibitemShut
  {NoStop}%
\bibitem [{\citenamefont {Chiba}\ and\ \citenamefont
  {Kimura}(2015)}]{Chiba2015}%
  \BibitemOpen
  \bibfield  {author} {\bibinfo {author} {\bibfnamefont {Y.}~\bibnamefont
  {Chiba}}\ and\ \bibinfo {author} {\bibfnamefont {M.}~\bibnamefont {Kimura}},\
  }\href {\doibase 10.1103/PhysRevC.91.061302} {\bibfield  {journal} {\bibinfo
  {journal} {Physical Review C - Nuclear Physics}\ }\textbf {\bibinfo {volume}
  {91}},\ \bibinfo {pages} {061302} (\bibinfo {year} {2015})}\BibitemShut
  {NoStop}%
\bibitem [{\citenamefont {Yamada}\ and\ \citenamefont
  {Funaki}(2015)}]{Yamada2015}%
  \BibitemOpen
  \bibfield  {author} {\bibinfo {author} {\bibfnamefont {T.}~\bibnamefont
  {Yamada}}\ and\ \bibinfo {author} {\bibfnamefont {Y.}~\bibnamefont
  {Funaki}},\ }\href {\doibase 10.1103/PhysRevC.92.034326} {\bibfield
  {journal} {\bibinfo  {journal} {Physical Review C - Nuclear Physics}\
  }\textbf {\bibinfo {volume} {92}},\ \bibinfo {pages} {034326} (\bibinfo
  {year} {2015})}\BibitemShut {NoStop}%
\bibitem [{\citenamefont {Chiba}\ \emph {et~al.}(2016)\citenamefont {Chiba},
  \citenamefont {Kimura},\ and\ \citenamefont {Taniguchi}}]{Chiba2016}%
  \BibitemOpen
  \bibfield  {author} {\bibinfo {author} {\bibfnamefont {Y.}~\bibnamefont
  {Chiba}}, \bibinfo {author} {\bibfnamefont {M.}~\bibnamefont {Kimura}}, \
  and\ \bibinfo {author} {\bibfnamefont {Y.}~\bibnamefont {Taniguchi}},\ }\href
  {\doibase 10.1103/PhysRevC.93.034319} {\bibfield  {journal} {\bibinfo
  {journal} {Physical Review C}\ }\textbf {\bibinfo {volume} {93}},\ \bibinfo
  {pages} {034319} (\bibinfo {year} {2016})}\BibitemShut {NoStop}%
\bibitem [{\citenamefont {Kanada-En'yo}(2016)}]{Kanada-EnYo2016a}%
  \BibitemOpen
  \bibfield  {author} {\bibinfo {author} {\bibfnamefont {Y.}~\bibnamefont
  {Kanada-En'yo}},\ }\href {\doibase 10.1103/PhysRevC.93.054307} {\bibfield
  {journal} {\bibinfo  {journal} {Physical Review C}\ }\textbf {\bibinfo
  {volume} {93}},\ \bibinfo {pages} {054307} (\bibinfo {year}
  {2016})}\BibitemShut {NoStop}%
\bibitem [{\citenamefont {Zhou}\ \emph {et~al.}(2016)\citenamefont {Zhou},
  \citenamefont {Tohsaki}, \citenamefont {Horiuchi},\ and\ \citenamefont
  {Ren}}]{Zhou2016}%
  \BibitemOpen
  \bibfield  {author} {\bibinfo {author} {\bibfnamefont {B.}~\bibnamefont
  {Zhou}}, \bibinfo {author} {\bibfnamefont {A.}~\bibnamefont {Tohsaki}},
  \bibinfo {author} {\bibfnamefont {H.}~\bibnamefont {Horiuchi}}, \ and\
  \bibinfo {author} {\bibfnamefont {Z.}~\bibnamefont {Ren}},\ }\href {\doibase
  10.1103/PhysRevC.94.044319} {\bibfield  {journal} {\bibinfo  {journal}
  {Physical Review C}\ }\textbf {\bibinfo {volume} {94}},\ \bibinfo {pages}
  {044319} (\bibinfo {year} {2016})}\BibitemShut {NoStop}%
\bibitem [{\citenamefont {Chiba}\ \emph {et~al.}(2017)\citenamefont {Chiba},
  \citenamefont {Taniguchi},\ and\ \citenamefont {Kimura}}]{Chiba2017b}%
  \BibitemOpen
  \bibfield  {author} {\bibinfo {author} {\bibfnamefont {Y.}~\bibnamefont
  {Chiba}}, \bibinfo {author} {\bibfnamefont {Y.}~\bibnamefont {Taniguchi}}, \
  and\ \bibinfo {author} {\bibfnamefont {M.}~\bibnamefont {Kimura}},\ }\href
  {\doibase 10.1103/PhysRevC.95.044328} {\bibfield  {journal} {\bibinfo
  {journal} {Physical Review C}\ }\textbf {\bibinfo {volume} {95}},\ \bibinfo
  {pages} {044328} (\bibinfo {year} {2017})},\ \Eprint
  {http://arxiv.org/abs/1610.04000} {arXiv:1610.04000} \BibitemShut {NoStop}%
\bibitem [{\citenamefont {Kanada-En'Yo}\ and\ \citenamefont
  {Shikata}(2019)}]{Kanada-EnYo2019}%
  \BibitemOpen
  \bibfield  {author} {\bibinfo {author} {\bibfnamefont {Y.}~\bibnamefont
  {Kanada-En'Yo}}\ and\ \bibinfo {author} {\bibfnamefont {Y.}~\bibnamefont
  {Shikata}},\ }\href {\doibase 10.1103/PhysRevC.100.014301} {\bibfield
  {journal} {\bibinfo  {journal} {Physical Review C}\ }\textbf {\bibinfo
  {volume} {100}},\ \bibinfo {pages} {014301} (\bibinfo {year}
  {2019})}\BibitemShut {NoStop}%
\bibitem [{\citenamefont {Chiba}\ and\ \citenamefont
  {Kimura}(2020)}]{Chiba2020}%
  \BibitemOpen
  \bibfield  {author} {\bibinfo {author} {\bibfnamefont {Y.}~\bibnamefont
  {Chiba}}\ and\ \bibinfo {author} {\bibfnamefont {M.}~\bibnamefont {Kimura}},\
  }\href {\doibase 10.1103/physrevc.101.024317} {\bibfield  {journal} {\bibinfo
   {journal} {Physical Review C}\ }\textbf {\bibinfo {volume} {101}},\ \bibinfo
  {pages} {024317} (\bibinfo {year} {2020})}\BibitemShut {NoStop}%
\bibitem [{\citenamefont {Kanada-En'yo}\ and\ \citenamefont
  {Ogata}(2020)}]{Kanada-EnYo2020}%
  \BibitemOpen
  \bibfield  {author} {\bibinfo {author} {\bibfnamefont {Y.}~\bibnamefont
  {Kanada-En'yo}}\ and\ \bibinfo {author} {\bibfnamefont {K.}~\bibnamefont
  {Ogata}},\ }\href {\doibase 10.1103/physrevc.101.014317} {\bibfield
  {journal} {\bibinfo  {journal} {Physical Review C}\ }\textbf {\bibinfo
  {volume} {101}},\ \bibinfo {pages} {014317} (\bibinfo {year}
  {2020})}\BibitemShut {NoStop}%
\bibitem [{\citenamefont {Berger}\ \emph {et~al.}(1991)\citenamefont {Berger},
  \citenamefont {Girod},\ and\ \citenamefont {Gogny}}]{Berger1991}%
  \BibitemOpen
  \bibfield  {author} {\bibinfo {author} {\bibfnamefont {J.}~\bibnamefont
  {Berger}}, \bibinfo {author} {\bibfnamefont {M.}~\bibnamefont {Girod}}, \
  and\ \bibinfo {author} {\bibfnamefont {D.}~\bibnamefont {Gogny}},\ }\href
  {\doibase 10.1016/0010-4655(91)90263-K} {\bibfield  {journal} {\bibinfo
  {journal} {Computer Physics Communications}\ }\textbf {\bibinfo {volume}
  {63}},\ \bibinfo {pages} {365} (\bibinfo {year} {1991})}\BibitemShut
  {NoStop}%
\bibitem [{\citenamefont {Kanada-En'yo}\ \emph {et~al.}(2003)\citenamefont
  {Kanada-En'yo}, \citenamefont {Kimura},\ and\ \citenamefont
  {Horiuchi}}]{Kanada-Enyo2003}%
  \BibitemOpen
  \bibfield  {author} {\bibinfo {author} {\bibfnamefont {Y.}~\bibnamefont
  {Kanada-En'yo}}, \bibinfo {author} {\bibfnamefont {M.}~\bibnamefont
  {Kimura}}, \ and\ \bibinfo {author} {\bibfnamefont {H.}~\bibnamefont
  {Horiuchi}},\ }\href {\doibase 10.1016/S1631-0705(03)00062-8} {\bibfield
  {journal} {\bibinfo  {journal} {Comptes Rendus Physique}\ }\textbf {\bibinfo
  {volume} {4}},\ \bibinfo {pages} {497} (\bibinfo {year} {2003})}\BibitemShut
  {NoStop}%
\bibitem [{\citenamefont {Kimura}(2004)}]{Kimura2004a}%
  \BibitemOpen
  \bibfield  {author} {\bibinfo {author} {\bibfnamefont {M.}~\bibnamefont
  {Kimura}},\ }\href {\doibase 10.1103/PhysRevC.69.044319} {\bibfield
  {journal} {\bibinfo  {journal} {Physical Review C}\ }\textbf {\bibinfo
  {volume} {69}},\ \bibinfo {pages} {044319} (\bibinfo {year}
  {2004})}\BibitemShut {NoStop}%
\bibitem [{\citenamefont {Kanada-En'yo}\ \emph {et~al.}(2012)\citenamefont
  {Kanada-En'yo}, \citenamefont {Kimura},\ and\ \citenamefont
  {Ono}}]{Kanada-Enyo2012}%
  \BibitemOpen
  \bibfield  {author} {\bibinfo {author} {\bibfnamefont {Y.}~\bibnamefont
  {Kanada-En'yo}}, \bibinfo {author} {\bibfnamefont {M.}~\bibnamefont
  {Kimura}}, \ and\ \bibinfo {author} {\bibfnamefont {A.}~\bibnamefont {Ono}},\
  }\href {\doibase 10.1093/ptep/pts001} {\bibfield  {journal} {\bibinfo
  {journal} {Progress of Theoretical and Experimental Physics}\ }\textbf
  {\bibinfo {volume} {2012}},\ \bibinfo {pages} {1A202} (\bibinfo {year}
  {2012})}\BibitemShut {NoStop}%
\bibitem [{\citenamefont {Baba}\ and\ \citenamefont {Kimura}(2016)}]{Baba2016}%
  \BibitemOpen
  \bibfield  {author} {\bibinfo {author} {\bibfnamefont {T.}~\bibnamefont
  {Baba}}\ and\ \bibinfo {author} {\bibfnamefont {M.}~\bibnamefont {Kimura}},\
  }\href {\doibase 10.1103/PhysRevC.94.044303} {\bibfield  {journal} {\bibinfo
  {journal} {Physical Review C}\ }\textbf {\bibinfo {volume} {94}},\ \bibinfo
  {pages} {044303} (\bibinfo {year} {2016})}\BibitemShut {NoStop}%
\bibitem [{\citenamefont {Baba}\ and\ \citenamefont {Kimura}(2017)}]{Baba2017}%
  \BibitemOpen
  \bibfield  {author} {\bibinfo {author} {\bibfnamefont {T.}~\bibnamefont
  {Baba}}\ and\ \bibinfo {author} {\bibfnamefont {M.}~\bibnamefont {Kimura}},\
  }\href {\doibase 10.1103/PhysRevC.95.064318} {\bibfield  {journal} {\bibinfo
  {journal} {Physical Review C}\ }\textbf {\bibinfo {volume} {95}} (\bibinfo
  {year} {2017}),\ 10.1103/PhysRevC.95.064318}\BibitemShut {NoStop}%
\bibitem [{\citenamefont {Hill}\ and\ \citenamefont
  {Wheeler}(1953)}]{Hill1953}%
  \BibitemOpen
  \bibfield  {author} {\bibinfo {author} {\bibfnamefont {D.~L.}\ \bibnamefont
  {Hill}}\ and\ \bibinfo {author} {\bibfnamefont {J.~A.}\ \bibnamefont
  {Wheeler}},\ }\href {\doibase 10.1103/PhysRev.89.1102} {\bibfield  {journal}
  {\bibinfo  {journal} {Physical Review}\ }\textbf {\bibinfo {volume} {89}},\
  \bibinfo {pages} {1102} (\bibinfo {year} {1953})}\BibitemShut {NoStop}%
\bibitem [{\citenamefont {Chiba}\ and\ \citenamefont
  {Kimura}(2017)}]{Chiba2017}%
  \BibitemOpen
  \bibfield  {author} {\bibinfo {author} {\bibfnamefont {Y.}~\bibnamefont
  {Chiba}}\ and\ \bibinfo {author} {\bibfnamefont {M.}~\bibnamefont {Kimura}},\
  }\href {\doibase 10.1093/ptep/ptx063} {\bibfield  {journal} {\bibinfo
  {journal} {Progress of Theoretical and Experimental Physics}\ }\textbf
  {\bibinfo {volume} {2017}} (\bibinfo {year} {2017}),\
  10.1093/ptep/ptx063}\BibitemShut {NoStop}%
\bibitem [{\citenamefont {Itagaki}\ and\ \citenamefont
  {Okabe}(2000)}]{Itagaki2000}%
  \BibitemOpen
  \bibfield  {author} {\bibinfo {author} {\bibfnamefont {N.}~\bibnamefont
  {Itagaki}}\ and\ \bibinfo {author} {\bibfnamefont {S.}~\bibnamefont
  {Okabe}},\ }\href {\doibase 10.1103/PhysRevC.61.044306} {\bibfield  {journal}
  {\bibinfo  {journal} {Physical Review C}\ }\textbf {\bibinfo {volume} {61}},\
  \bibinfo {pages} {044306} (\bibinfo {year} {2000})}\BibitemShut {NoStop}%
\bibitem [{\citenamefont {von Oertzen}\ \emph {et~al.}(2006)\citenamefont {von
  Oertzen}, \citenamefont {Freer},\ and\ \citenamefont
  {Kanada-En'yo}}]{VonOertzen2006}%
  \BibitemOpen
  \bibfield  {author} {\bibinfo {author} {\bibfnamefont {W.}~\bibnamefont {von
  Oertzen}}, \bibinfo {author} {\bibfnamefont {M.}~\bibnamefont {Freer}}, \
  and\ \bibinfo {author} {\bibfnamefont {Y.}~\bibnamefont {Kanada-En'yo}},\
  }\href {\doibase 10.1016/j.physrep.2006.07.001} {\bibfield  {journal}
  {\bibinfo  {journal} {Physics Reports}\ }\textbf {\bibinfo {volume} {432}},\
  \bibinfo {pages} {43} (\bibinfo {year} {2006})}\BibitemShut {NoStop}%
\bibitem [{\citenamefont {Kimura}(2007)}]{Kimura2007}%
  \BibitemOpen
  \bibfield  {author} {\bibinfo {author} {\bibfnamefont {M.}~\bibnamefont
  {Kimura}},\ }\href {\doibase 10.1103/PhysRevC.75.034312} {\bibfield
  {journal} {\bibinfo  {journal} {Physical Review C - Nuclear Physics}\
  }\textbf {\bibinfo {volume} {75}} (\bibinfo {year} {2007}),\
  10.1103/PhysRevC.75.034312}\BibitemShut {NoStop}%
\bibitem [{\citenamefont {Baba}\ \emph {et~al.}(2014)\citenamefont {Baba},
  \citenamefont {Chiba},\ and\ \citenamefont {Kimura}}]{Baba2014}%
  \BibitemOpen
  \bibfield  {author} {\bibinfo {author} {\bibfnamefont {T.}~\bibnamefont
  {Baba}}, \bibinfo {author} {\bibfnamefont {Y.}~\bibnamefont {Chiba}}, \ and\
  \bibinfo {author} {\bibfnamefont {M.}~\bibnamefont {Kimura}},\ }\href
  {\doibase 10.1103/PhysRevC.90.064319} {\bibfield  {journal} {\bibinfo
  {journal} {Physical Review C}\ }\textbf {\bibinfo {volume} {90}},\ \bibinfo
  {pages} {064319} (\bibinfo {year} {2014})}\BibitemShut {NoStop}%
\bibitem [{\citenamefont {Li}\ \emph {et~al.}(2017)\citenamefont {Li},
  \citenamefont {Ye}, \citenamefont {Li}, \citenamefont {Lin}, \citenamefont
  {Li}, \citenamefont {Ge}, \citenamefont {Lou}, \citenamefont {Tian},
  \citenamefont {Jiang}, \citenamefont {Yang}, \citenamefont {Feng},
  \citenamefont {Li}, \citenamefont {Chen}, \citenamefont {Liu}, \citenamefont
  {Zang}, \citenamefont {Yang}, \citenamefont {Zhang}, \citenamefont {Chen},
  \citenamefont {Liu}, \citenamefont {Sun}, \citenamefont {Ma}, \citenamefont
  {Jia}, \citenamefont {Xu}, \citenamefont {Yang}, \citenamefont {Ma},\ and\
  \citenamefont {Sun}}]{Li2017}%
  \BibitemOpen
  \bibfield  {author} {\bibinfo {author} {\bibfnamefont {J.}~\bibnamefont
  {Li}}, \bibinfo {author} {\bibfnamefont {Y.~L.}\ \bibnamefont {Ye}}, \bibinfo
  {author} {\bibfnamefont {Z.~H.}\ \bibnamefont {Li}}, \bibinfo {author}
  {\bibfnamefont {C.~J.}\ \bibnamefont {Lin}}, \bibinfo {author} {\bibfnamefont
  {Q.~T.}\ \bibnamefont {Li}}, \bibinfo {author} {\bibfnamefont {Y.~C.}\
  \bibnamefont {Ge}}, \bibinfo {author} {\bibfnamefont {J.~L.}\ \bibnamefont
  {Lou}}, \bibinfo {author} {\bibfnamefont {Z.~Y.}\ \bibnamefont {Tian}},
  \bibinfo {author} {\bibfnamefont {W.}~\bibnamefont {Jiang}}, \bibinfo
  {author} {\bibfnamefont {Z.~H.}\ \bibnamefont {Yang}}, \bibinfo {author}
  {\bibfnamefont {J.}~\bibnamefont {Feng}}, \bibinfo {author} {\bibfnamefont
  {P.~J.}\ \bibnamefont {Li}}, \bibinfo {author} {\bibfnamefont
  {J.}~\bibnamefont {Chen}}, \bibinfo {author} {\bibfnamefont {Q.}~\bibnamefont
  {Liu}}, \bibinfo {author} {\bibfnamefont {H.~L.}\ \bibnamefont {Zang}},
  \bibinfo {author} {\bibfnamefont {B.}~\bibnamefont {Yang}}, \bibinfo {author}
  {\bibfnamefont {Y.}~\bibnamefont {Zhang}}, \bibinfo {author} {\bibfnamefont
  {Z.~Q.}\ \bibnamefont {Chen}}, \bibinfo {author} {\bibfnamefont
  {Y.}~\bibnamefont {Liu}}, \bibinfo {author} {\bibfnamefont {X.~H.}\
  \bibnamefont {Sun}}, \bibinfo {author} {\bibfnamefont {J.}~\bibnamefont
  {Ma}}, \bibinfo {author} {\bibfnamefont {H.~M.}\ \bibnamefont {Jia}},
  \bibinfo {author} {\bibfnamefont {X.~X.}\ \bibnamefont {Xu}}, \bibinfo
  {author} {\bibfnamefont {L.}~\bibnamefont {Yang}}, \bibinfo {author}
  {\bibfnamefont {N.~R.}\ \bibnamefont {Ma}}, \ and\ \bibinfo {author}
  {\bibfnamefont {L.~J.}\ \bibnamefont {Sun}},\ }\href {\doibase
  10.1103/PhysRevC.95.021303} {\bibfield  {journal} {\bibinfo  {journal}
  {Physical Review C}\ }\textbf {\bibinfo {volume} {95}},\ \bibinfo {pages}
  {021303} (\bibinfo {year} {2017})}\BibitemShut {NoStop}%
\bibitem [{\citenamefont {Ikeda}\ \emph {et~al.}(1968)\citenamefont {Ikeda},
  \citenamefont {Takigawa},\ and\ \citenamefont {Horiuchi}}]{Ikeda1968}%
  \BibitemOpen
  \bibfield  {author} {\bibinfo {author} {\bibfnamefont {K.}~\bibnamefont
  {Ikeda}}, \bibinfo {author} {\bibfnamefont {N.}~\bibnamefont {Takigawa}}, \
  and\ \bibinfo {author} {\bibfnamefont {H.}~\bibnamefont {Horiuchi}},\ }\href
  {\doibase 10.1143/PTPS.E68.464} {\bibfield  {journal} {\bibinfo  {journal}
  {Progress of Theoretical Physics Supplement}\ }\textbf {\bibinfo {volume}
  {E68}},\ \bibinfo {pages} {464} (\bibinfo {year} {1968})}\BibitemShut
  {NoStop}%
\end{thebibliography}%
\end{document}